\documentclass[superscriptaddress,aps,pra,twocolumn,showpacs,nofootinbib,longbibliography,floatfix]{revtex4-2}

\usepackage[T1]{fontenc}
\usepackage{amsfonts}
\usepackage{amsmath,amssymb,amsthm,amstext,mathrsfs}
\usepackage{latexsym}
\usepackage{physics}
\usepackage{gensymb}
\usepackage{braket}

\usepackage{graphicx}
\usepackage{subcaption}
\usepackage{float}
\usepackage{adjustbox}
\usepackage{booktabs}
\usepackage{color}
\usepackage{xcolor}
\usepackage[colorlinks=true,citecolor=blue,urlcolor=blue]{hyperref}

\usepackage{soul}        
\usepackage[normalem]{ulem} 
\usepackage{lipsum}

\newcommand{\be}{\begin{equation}}
\newcommand{\ee}{\end{equation}}
\newcommand{\ba}{\begin{eqnarray}}
\newcommand{\ea}{\end{eqnarray}}

\begin{document}
	
\title{Robust certification of quantum instruments through a sequential  communication game}

\author{Pritam Roy}
\email{roy.pritamphy@gmail.com}
\affiliation{S. N. Bose National Centre for Basic Sciences, Block JD, Sector III, Salt Lake, Kolkata 700 106, India}

\author{Subhankar Bera}
\email{berasanu007@gmail.com}
\affiliation{S. N. Bose National Centre for Basic Sciences, Block JD, Sector III, Salt Lake, Kolkata 700 106, India}

\author{A. S. Majumdar}
\email{archan@bose.res.in}
\affiliation{S. N. Bose National Centre for Basic Sciences, Block JD, Sector III, Salt Lake, Kolkata 700 106, India}

\author{Shiladitya Mal}
\email{shiladitya.27@gmail.com}
\affiliation{Centre for Quantum Science and Technology, Chennai Institute of Technology, Chennai, Tamil Nadu 600 069, India}

\begin{abstract}
We propose a  communication game in the sequential measurement scenario,
involving a sender and two receivers with restricted communication
among the latter parties. In the framework of the prepare-transform-measure scenario, we find a prominent quantum advantage in the receiver's decoding of the message originally encoded by the sender. We show that an optimal trade-off between the success probabilities of the two receivers enables self-testing of the sender's state preparation, the first receiver's instruments,
and the measurement device of the second receiver in a semi device-independent way. Our
protocol enables a more robust certification of the unsharp measurement parameter of the first receiver compared to an earlier protocol. We further generalize our game
to higher-dimensional systems, revealing greater quantum advantage with
an increase in dimensions.

\end{abstract}

\maketitle

\section{Introduction} 
The study of the performance of quantum systems for communication and computational purposes gave birth to quantum information science \cite{nielsen_chuang_2010, Watruos_QisBook}. The study was essential as the technology was maturing towards manipulating single quantum systems \cite{nobel_12}. In present times, the certification of quantum systems, is regarded as an important cornerstone for the successful and secure implementation of quantum technology protocols. In the literature, various types of certification schemes are classified according to the set of assumptions and level of trust required for the experimental implementation. The certification scheme is called robust if it remains meaningful in presence of some amount of deviation from the ideal description. 
Robust self-testing usually refers to the certification of quantum states and measurements in a device-independent manner \cite{POPESCU1992411, self-testing_MY_04, Bancal_ST_15, chen_st_16, kaniewski_st_17, Supic2020selftestingof, mal_quantboundary_23}. In semi-device independent scenarios, requirements are lesser stringent, which is practically more relevant \cite{vsupic2016self, bowles2014one, quintino2015inequivalence, wollmann2016observation, goswami2018one, bian2020experimental}.  Witnesses which require trust on preparation and measurement devices, belong to the category of device-dependent formalism \cite{Horodecki_RMP09, GUHNE2009}.

Historically, considering the communication capacity of a photon as a carrier of classical information, Holevo derived a crucial result \cite{Holevo73}, which implies that a qubit cannot carry more than a bit of information. This apparent conclusion that a qubit is not more powerful than a bit was belied later due to a clever protocol that can reveal the power of a qubit over a bit. Specifically, the communication task of Random Access Codes (RAC) was developed on the conjugate coding protocol \cite{wisener_83}, in the prepare-and-measure scenario, which exhibits the supremacy of a qubit over a bit \cite{Ambainis_DCQA99, Ambainis_DCFA02, Ambainis_RACSR09}. RAC is a fundamental protocol in information theory. In the case of classical $2\rightarrow 1$ RAC, two bits of classical information are compressed into a bit and are sent to a receiver, who has to retrieve any one of the bit values when asked randomly. The success probability of retrieving the encoded message is the figure of merit of the game.
On the other hand, in the quantum version of random access code (QRAC), the encoding is done on a quantum system. It was shown that the success probability of correctly guessing a bit is larger when a qubit is a carrier than a bit in $2\rightarrow 1$ the RAC protocol \cite{Ambainis_DCQA99, Ambainis_DCFA02}. Later, quantum advantage has also been demonstrated for higher-dimensional QRAC \cite{Tavakoli_QRAC15}. 

QRAC has emerged as a basic building block in various quantum information processing tasks, such as semi-device-independent quantum key distribution \cite{Pawlowski_SDI11}, dimension witness \cite{Brunner_DW13, Tavakoli_CEPM21, Wehner_DW08}, randomness generation \cite{Li_SDRNG11, Li_SDRNG12}, quantum bidding  \cite{qBridge_exp_14}, witnessing incompatibility of measurement \cite{saha_incompatible_23}. 
The prepare and measure (PM) scenario enables 
certification of the quantum state and measurements  \cite{Tavakoli_STPM18, Mironowicz_SDIM19, Maity_st_21, das2022robust} in a semi-device-independent way, while device-independent self-testing \cite{POPESCU1992411, self-testing_MY_04, Bancal_ST_15, chen_st_16, kaniewski_st_17, Supic2020selftestingof, mal_quantboundary_23}  relies on the Bell test \cite{Bell_64, chsh_69}. In addition to various quantum information processing tasks, QRAC has been investigated in association with foundational issues of quantum mechanics as well \cite{spekkens_pc_09, Pawlowski_IC_09, bera_QRAC_22, Mal_FURPC_21, gupta2023quantum}. Various experimental 
manifestations based on QRAC protocols have been proposed  \cite{spekkens_pc_09, qBridge_exp_14, Aguilar_pqrac_18, Tavakoli_QRAC15, foletto_ExpSQRAC_20, anwer_expSQRAC_20, xiao_expSQRAC_21}.

On the other hand, the idea of employing sequential measurements in investigating foundational and practical information processing issues have gained much importance in recent years \cite{silva_SequentialBell_15, mal_SequentialBell_16, mal_sequentialSteering_18, mal_sequentialEntaglement_18, das2019facets, Akshata_19, Asmita_19, brown2020arbitrarily, cheng2021limitations, mal_SequentialCGLMP_24, munshi2025device}. The sequential Bell test scenario has been employed to harness unbounded randomness from a single pair of entangled states \cite{Curchod_UNrand_17} and self-testing quantum instruments as well \cite{Wagner_STQInstruments_20}. Moreover, sequential measurements have been employed to study various other quantum correlations such as quantum steering \cite{datta2018sharing, maity2020detection,gupta2021genuine}, and sequential communication protocols have been suggested for information processing tasks such as entanglement detection, teleportation,  and remote state preparation \cite{das2022resource, roy2021recycling, datta2024remote}. Here we refer to the non-sequential setting as the standard one where certification of quantum instruments, non-projective and unsharp measurements is not possible. This can be circumvented invoking sequential settings \cite{Mohan_STI19, Miklin_SDIM20, Tavakoli_STM20}.  The certification of quantum instruments, which plays a key role in quantum science and technology,  is the main focus of the present paper.

The generalization of the standard $2\rightarrow 1$ QRAC protocol to the
sequential scenario has  been performed \cite{ Mohan_STI19, Miklin_SDIM20, debarshi_seq_RAC24}, enabling various certification tasks. In \cite{Mohan_STI19}, $2\rightarrow 1$ QRAC was extended to three parties where two independent receivers try to decode the bit position encoded earlier by the sender, randomly and sequentially. Here, the first decoder must have to measure weakly so that the second decoder can have a non-classical success probability. In this prepare-transform-measure (PTM) scenario, the success probability of the two decoders exhibits a trade-off, absent in the analogous classical counterpart. Such a protocol can be employed to self-test quantum instruments \cite{Mohan_STI19}, and non-projective \cite{Tavakoli_STM20} and unsharp measurements \cite{Miklin_SDIM20} in a semi-device-independent manner going beyond the standard settings of certification of quantum states and measurements. It may be noted that experimental implementations of  sequential measurement
protocols have also been performed \cite{schiavon2017three,hu2018observation}.

In the present work, we propose a new type of communication game and show its quantum advantage, which enables certification of various quantum components in a semi device-independent way. Surprisingly, our scheme yields a more robust certification of quantum instruments, known so far. The standard $2\rightarrow1$ QRAC scenario is extended to a tripartite communication game where there are two sequential receivers who are not completely independent, unlike the previous extension \cite{Mohan_STI19}. Here, the sender (Aparna) and the first receiver (Barun) act in the same way as in the standard RAC task, whereas the second receiver (Chhanda) has to guess the remaining bit position so that at the end of the game, both receivers together may have the full information about the message encoded by the sender. Specifically, in the context of $2\rightarrow1$ QRAC, two bits of information are encoded into one bit by Aparna. Barun decodes the bit value of some position according to his random input. After decoding the asked bit value, Barun passes the post-measured system to Chhanda without leaking any output information, and the task of Chhanda is to guess the other bit value. We compute the optimal classical success probability of this communication game and demonstrate a prominent advantage through a quantum strategy. 

Next, we explore a trade-off between the success probability of Barun and Chhanda, which is absent in the classical case. The optimal trade-off leads to the certification of Barun's quantum instrument together with Aparna's state preparation and Chhanda's measurement. We then investigate the robustness of the certification of quantum instruments in the presence of noise and estimate the lower and upper bounds on the sharpness parameter of Barun's measurement. A significant outcome of our analysis is that our certification task is shown to be more robust compared to the approach of employing sequential QRAC without collaboration \cite{Mohan_STI19}. Since the gap between the lower and upper bounds of the estimated sharpness parameter of the first receiver's measurement is narrower in our case, we are able to improve upon the earlier results in the context of our new communication game. We further generalize our game to the higher-dimensional $2^{d-1}\rightarrow 1$ QRAC scenario. We find the optimal classical success probability and numerically compute the quantum advantage up to dimension six. 


The manuscript is organized in the following manner. We begin with the description of the new communication game in the next section. In Section III, we derive a trade-off relation between
the success probabilities of the two receivers and show how this enables
robust certification of the quantum components. Next, we generalize the proposed communication task to the higher-dimensional case in Section IV.  We conclude with a summary of our analysis and results in Section V. 

\section{Two receiver sequential game with restricted communication} 

\begin{figure}[htbp]
\begin{center}
\includegraphics[width=8.5cm, keepaspectratio]{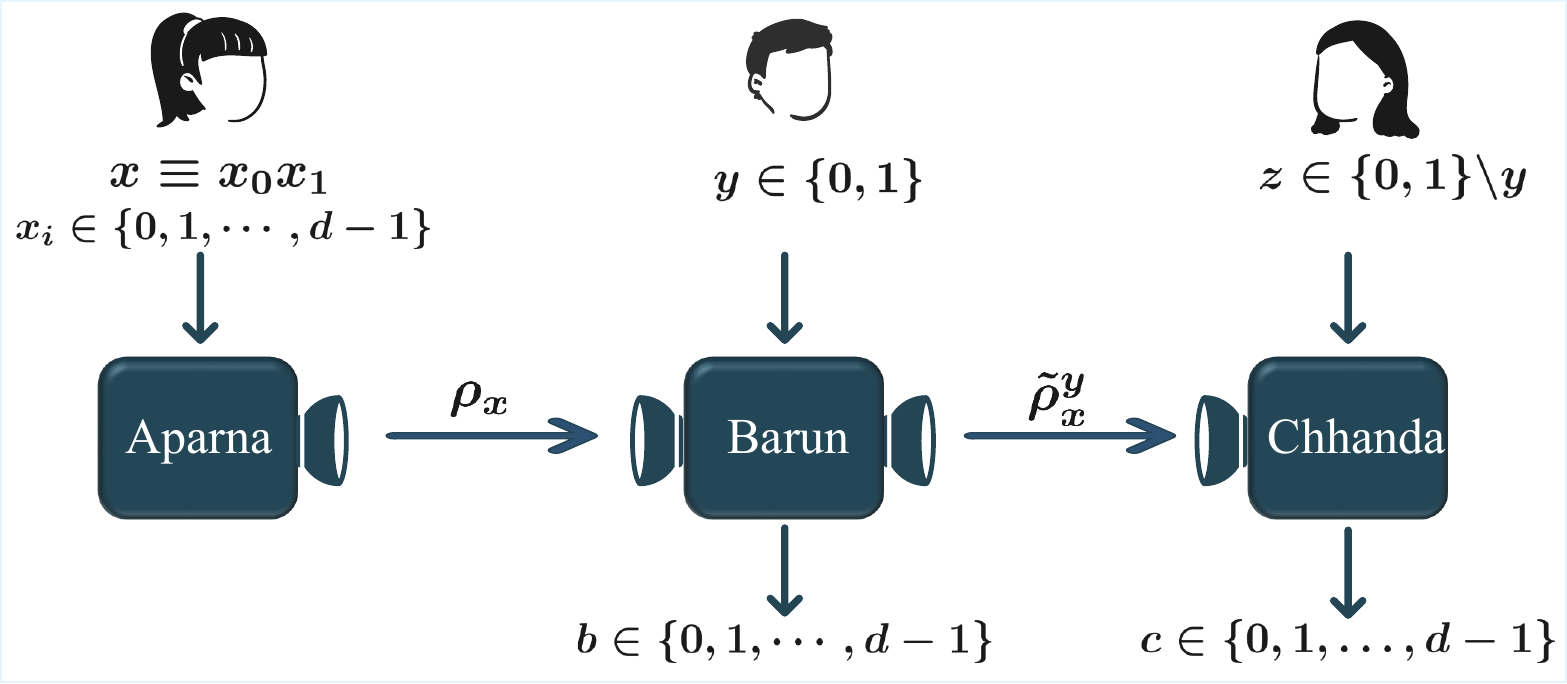}
\caption{\footnotesize Schematic diagram for a two-receiver communication game with restricted collaboration in `d'-dimensional quantum settings. Sender Aparna compressed a 2-dit message into a qudit and sends it to the first receiver, Barun, who decodes a dit position randomly. The task of the second receiver, Chhanda, is to guess the leftover information under the restriction that she does not have Barun's output information except wild guess.}
\label{modrac}
\end{center}
\end{figure}

In this section, we describe our proposed new communication game pertaining to the prepare-transform and measure scenario among three parties: Aparna, Barun, and Chhanda. Initially, two dits (higher-dimensional generalizations of a bit) of information are encoded into one dit by the sender, Aparna. The first receiver, Barun, tries to decode some partial message according to the random input, obtained from some referee, by measuring the encoded system and then passes it to Chhanda, who tries to retrieve the remaining message. The crucial point is that, unlike the protocol \cite{Mohan_STI19}, here two receivers are not independent, and Barun's communication is such that Chhanda can not have his output information except wild guess. In the next two subsections, we will discuss both the classical and quantum strategies and relevant pay-off functions of this game, followed by results in the subsequent sections.

\subsection{Two-receiver classical access code}
\label{ClaG}
In the classical version of this game, Aparna first selects a message randomly from $d^2$ options, which is a 2-dit string \( x = (x_0, x_1) \). She encodes her two dits \((x_0, x_1) \in \{0,1,\dots,d-1\}^2\) into a one-dit message \(m \in \{0,1,\dots,d-1\}\) and sends it to Barun. Barun and Chhanda receive a common random input \(y \in \{0,1\}\), say, from a referee. Barun’s task is to guess the dit \(x_y\), while Chhanda’s task is to guess the complementary dit \(x_{\bar y}\). After producing his guess, Barun forwards the post-measurement system to Chhanda, satisfying the constraint that the system carries no information about his output. Then Chhanda attempts to infer \(x_{\bar y}\) (see Fig.~\ref{modrac}) using the system.

The performance of this communication task is evaluated via three payoff functions. They are quantified by the average success probability of correctly guessing the asked bit value individually by Barun, Chhanda and that of by them jointly. \\
\textbf{First payoff:} The average success probability of Barun is given by,
\begin{equation}
\label{BarunSuccessC}
\begin{aligned}
    P_{AB}^{C}(b=x_y|x,y)   &= \frac{1}{2 d^2} \sum_{x, y} p(b = x_y |x, y) 
\end{aligned}
\end{equation}

In this task, Barun’s role is similar to that of the traditional RAC \cite{Ambainis_RACSR09, Tavakoli_QRAC15}. To maximize Barun's success probability of correctly retrieving the asked dit value, they can adopt the first-dit encoding strategy without loss of generality. With this approach, when Barun’s input is \( y = 0 \), he retrieves the dit with probability 1. However, for the second dit (when \( y = 1 \)), his success probability is \( \frac{1}{d} \) as he has no option other than a random guess. Therefore, on average, his success probability (\(P_{AB}^C\)) is \( \frac{1}{2} \left(1 + \frac{1}{d}\right) \). This classical strategy was conjectured to be optimal \cite{Tavakoli_QRAC15} and proved recently in \cite{Ambainis_QRAC24}. 

\textbf{Second payoff:} Similarly, the average success probability of Chhanda is given by,
\begin{equation}
\label{ChhandaesuccessC}
    \begin{aligned}
    P_{AC}^{C}(c=x_{\bar{y}}|x,y) &= \frac{1}{2 d^2} \sum_{x, \bar{y}} p(c = x_{\bar{y}} |x, \bar{y}). 
\end{aligned}
\end{equation}

Unlike Barun, Chhanda can't do better than a wild guess. Since she receives only the post-measurement system from Barun which carries no information about his output. Thus, her success probability \((P_{AC}^C)\) is limited to \(\tfrac{1}{d}\). 

\textbf{Third payoff:} The total classical success probability of retrieving the sender's full encoded message, jointly by the two receivers, according to the above stated strategy of restricted collaboration, is given by, 
\begin{equation}\label{cl_suc_prob}
    P^C(b=x_y,c=x_{\bar{y}}|x,y) = \frac{1}{2d} \left(1 + \frac{1}{d}\right).
\end{equation}

\subsection{Two-receiver quantum access code}\label{QG}
Now let's describe the two receiver quantum access code i.e., quantum version of the proposed game. Aparna prepares a quantum state $\rho_x$ that encodes the selected classical message, \( x = (x_0, x_1) \) $\in \{0, 1, ...d-1\}^2$. Two receivers, Barun and Chhanda, only know the set of states of dimension \( d \) from which Aparna selects in a single run of the game, but not the exact state. After preparing the state, Aparna sends it to Barun, who applies his measuring instrument, represented mathematically by the Kraus operators \( \{K_{b|y}\} \). When this instrument acts on the state \( \rho_x \), the outcome is \( b \in \{0, 1, \dots, d-1\} \), and the updated state according to the L\"udder transformation rule \cite{Busch1986}, is given by,

\begin{equation}
\rho_{x}^{y,b} = \frac{K_{b|y} \rho_{x} K_{b|y}^\dagger}{\text{Tr}(\rho_{x} K_{b|y} K_{b|y}^\dagger)}.
\end{equation}

To satisfy the completeness relation for Barun's measurement, we require that \( \sum_{b=0}^{d-1} M_{b|y} = \mathbb{I} \), where \( M_{b|y} = K_{b|y}^\dagger K_{b|y} \) are the elements of a positive operator-valued measure (POVM). The correlation between Aparna and Barun is captured by the average success probability of correctly guessing the asked dit value, which constitute the \emph{first payoff} and is given by, 
\begin{equation}
\label{BarunSuccess}
\begin{aligned}
    P_{AB}^{Q}(b=x_y|x,y)   &= \frac{1}{2 d^2} \sum_{x, y} p(b = x_y |x, y) \\  &= \frac{1}{2 d^2} \sum_{x, y} Tr(\rho_x M_{x_{y}|y})
\end{aligned}
\end{equation}
In the present work, we use the words average success probability, success probability, and figure of merit of the game interchangeably. 
According to the rules of the game, Barun only passes his post-measured state without leaking output information. Therefore, the post-measurement state received by Chhanda, is given by the average over all possible outcomes for a fixed input \( y \), which is,
\begin{equation}\label{av_state}
    \Tilde{\rho}_{x}^{y} = \sum_{b} K_{b|y}\rho_x K_{b|y}^{\dagger}.
\end{equation}
Now, to retrieve the leftover information, Chhanda will perform a generalized measurement $\{N_0, N_1\}$ on $\Tilde{\rho}_{x}^{y}$, where, $\sum_{b = 0}^{d-1}N_{b|y} = \mathbb{I}$. Chhanda's figure of merit is the success probability of correctly guessing the remaining part of the encoding, which is the \emph{second payoff function} of the game, and computed as given bellow.
\begin{equation}
\label{Chhandaesuccess}
    \begin{aligned}
    P_{AC}^{Q}(c=x_{\bar{y}}|x,y) &= \frac{1}{2 d^2} \sum_{x, \bar{y}} p(c = x_{\bar{y}} |x, \bar{y}) \\
           &= \frac{1}{2 d^2} \sum_{x, b, \bar{y}} \text{Tr}(K_{b|y}\rho_x K_{b|y}^{\dagger} N_{x_{\bar{y}}|\bar{y}}). 
\end{aligned}
\end{equation}
Our \emph{third payoff}, the total average success probability of retrieving the full information of Aparna's encoding by Barun and Chhanda together employing this restricted collaborative strategy is
\begin{equation}\label{q_joint_sucprob}
    P^{Q}(b=x_y,c=x_{\bar{y}}|x,y) = P_{AB}^{Q}P_{AC}^{Q}.
\end{equation}

If \(P_{AB}^{Q} > P_{AB}^{C}\) and \(P_{AC}^{Q} > P_{AC}^{C}\), then both Barun and Chhanda obtain a quantum advantage and similar conclusion holds for joint success probability when \(P^{Q} > P^{C}\) is obtained. 

\section{Quantum advantage and robust certification of instruments}\label{Result}
In this section, we analyze quantum strategy of the one-sender, two receiver new communication game. We show explicitly that in quantum settings, an interesting non-trivial trade-off between the first and the second payoff function, which leads to robust certification of quantum instruments along with SDI certification of states and measurements.  We consider the minimal scenario here, i.e., a two-bit message compressed into one bit, and generalize it to higher dimensions in the next section.

\subsection{Trade-off between Barun's and Chhanda's respective figure of merit}
Whatever strategy we follow, our goal is to maximize the success probability of the second receiver given that the first receiver obtains a fixed value of the non-classical success probability. This consideration characterizes the set of allowed regions in the correlation space. In the classical version of the game (Sec.~\ref{ClaG}), Barun and Chhanda's success probabilities are independent, resulting in a trivial scenario with no observable trade-off. In contrast, the quantum strategy (Sec.~\ref{QG}) introduces a more complex interplay between these probabilities. 

To formulate the problem in the quantum setting, we fix the value of \(P_{AB}^Q\) (denoted as \(\beta\)) and seek to determine the maximum achievable value of \(P_{AC}^Q\) over state preparations, instruments, and measurements under the relevant constraints by casting this problem as an SDP, which is explicitly given below. 
Our analysis focuses on the range \(\beta \in \left[\frac{1}{2}, \frac{1}{2}\left(1 + \frac{1}{\sqrt{2}}\right)\right]\), which encapsulates the optimal trade-off between the success probability of two receivers. This range represents the nontrivial boundary of the quantum set of correlations in the \((P_{AB}^Q, P_{AC}^Q)\) plane. It is to be mentioned here that, unlike the sequential QRAC \cite{Mohan_STI19}, here the second receiver has no random input, and together both Barun and Chhanda have a non-zero probability of decoding the full classical information encoded by Aparna. 

Formally, this leads to the following optimization problem: 

\begin{equation}
\begin{aligned}
P_{AC}^{Q,\beta} &= \max_{\rho, U, M, N} P_{AC}^Q \\
\text{s. t.} \quad 
&\forall x : \rho_x \in \mathbb{C}^2, \; \rho_x \geq 0, \; \text{Tr}(\rho_x) = 1, \\
&\forall z, c : N_{c|z} \geq 0, \; N_{0|z} + N_{1|z} = \mathbb{I}, \\
&\forall y, b : U_{yb} \in \mathrm{SU}(2),\\& \; M_{b|y} \geq 0, \; M_{0|y} + M_{1|y} = \mathbb{I}, \\
& \forall y, ~~\frac{1}{4}\sum_{x}\tilde{\rho}_x^y=\frac{\mathbb{I}}{2},\\
&\text{and } P_{AB}^{Q} = \beta.
\end{aligned}
\label{optimisation}
\end{equation}
\begin{figure}[ht]
\begin{center}
\includegraphics[width=8.5cm, keepaspectratio]{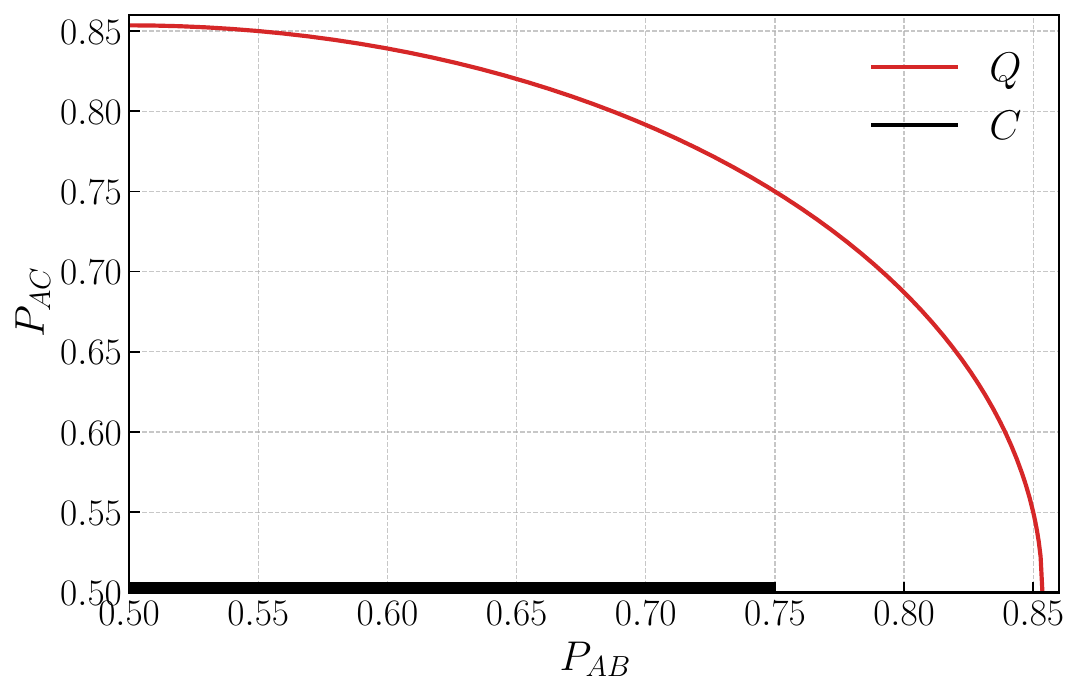}
\caption{\footnotesize Trade-off between the success probabilities of two receivers is shown. The region on and below the red line is achievable employing a quantum strategy. Therefore, the red line refers to the optimal trade-off and hence, the quantum boundary. The deep dark line indicates the classical boundary, beyond which is the non-classical region. }
\label{modrac2}
\end{center}
\end{figure}

In the above SDP~(\ref{optimisation}), we maximize Chhanda’s quantum success
probability by optimizing over all possible state preparations and measurement
strategies. The first constraint ensures that Aparna’s encodings \(\rho_x\) are
valid qubit states. The second and third constraints specify that both
Barun’s and Chhanda’s measurements are valid POVMs, with Barun’s instruments
further expressed in Kraus form
\(K_{b|y} = U_{y|b}\sqrt{M_{b|y}}\), where \(U_{y|b}\) is a unitary and
\(\{M_{b|y}\}\) a POVM. The condition \(P_{AB}^Q = \beta\) is imposed through
these instruments to fix Barun’s performance level. The second last constraint
\(\frac{1}{4}\sum_x \tilde{\rho}_x^y = \tfrac{\mathbb{I}}{2}\) enforces the restriction
that the system relayed by Barun contains no information about his output,
ensuring that Chhanda only has access to Barun’s input but not his outcome.

As the game between Aparna and Barun is similar to the standard QRAC \cite{Ambainis_RACSR09, Tavakoli_QRAC15}, we can employ a quantum strategy for  the above SDP (\ref{optimisation}), where Aparna  encodes
her preparations $\{\rho_{x}\}$ as,
\begin{equation}
\begin{aligned}
\rho_{00} &= \frac{1}{2}\left(\mathbb{I}+\frac{\sigma_x + \sigma_z}{\sqrt{2}}\right), \; \rho_{01} &= \frac{1}{2}\left(\mathbb{I}+\frac{\sigma_x - \sigma_z}{\sqrt{2}}\right),\\
\rho_{11} &= \frac{1}{2}\left(\mathbb{I}-\frac{\sigma_x + \sigma_z}{\sqrt{2}}\right), \; \rho_{10} &= \frac{1}{2}\left(\mathbb{I}-\frac{\sigma_x - \sigma_z}{\sqrt{2}}\right),
\end{aligned}
\label{enPrep}
\end{equation}
and Barun  performs an unsharp measurement, $M_0 = \eta \sigma_x$, $M_1 = \eta \sigma_z$ with sharpness parameter $\eta \in [0,1]$. Next, Chhanda  performs $N_0 = \sigma_x$, $N_1 = \sigma_z$ measurements to optimize her success probability. Using this specified strategy, their success probability is given by,
\begin{equation}
    \begin{aligned}
        P_{AB}^{Q} &= \frac{1}{2}(1+\frac{\eta}{\sqrt{2}})\\
        P_{AC}^{Q} &= \frac{1}{4}(2 + \sqrt{2-2 \eta^2})
    \end{aligned}
\label{tradeoff}
\end{equation}

The trade-off between $P_{AB}^{Q}$ and $P_{AC}^{Q}$ is displayed in the Fig.\ref{modrac2}. The above quantum strategy is actually optimal, since it saturates the upper bound on $P_{AC}^{Q}$ for any chosen value of $P_{AB}^{Q} = \beta$ (see Appendix $\eqref{proof_lemma}$). 
  Next, we move on to show explicitly how the optimal success probability point on the quantum boundary of the trade-off curve in correlation space leads to the certification of various quantum components involved in the strategy in a semi-device-independent way.

 
\subsection{Self-testing based on optimal quantum strategy} \label{selftest}
For the purpose of self-testing, we need to identify the underlying quantum objects uniquely up to some symmetry operations. The optimal trade-off relation allows us to certify Aparna's state preparation, Barun's instruments, and Chhanda's measurements within a collectively chosen reference frame. For more detailed proof of the optimal trade-off and self-testing argument, see the Appendix $\eqref{proof_lemma}$.  

An optimal pair of average quantum success probabilities of this multi-receiver quantum access protocol $(P_{AB}^Q, P_{AC}^Q)$ = $(\beta,P_{AC}^{\beta})$, as given by \eqref{tradeoff}, self-tests the following physical entities
\begin{itemize}
    \item Aparna’s state preparations are as given in Eq.\eqref{enPrep}, which are pure and form a square on the Bloch sphere representation.
    \item Barun's measurement instruments are given by the Kraus operators $ K_{b|y} = U_{y|b} \sqrt{M_{b|y}}$. The unitary freedom turns out to be the same for every observable and $M_0=\eta\sigma_x, M_1=\eta\sigma_z$, where $\eta=\sqrt{2}(2\beta - 1)$.
    \item Chhanda’s rank-1 projective measurements $\{N_0, N_1\}$ are given by $N_0=U\sigma_x U^{\dagger}, N_1=U\sigma_z U^{\dagger}$.
\end{itemize}

The optimal trade-off relation in the sequential protocol enables self-testing of crucial quantum components; i) Aparna’s square-shaped pure state preparations on the Bloch sphere, ii) Barun’s measurement instruments defined by Kraus operators with consistent unitary freedom, and iii) Chhanda’s rank-1 projective measurements equivalent to Pauli operators under a unitary transformation. \emph{It is to be noted that robust self-testing of states and measurements similar to i) and iii), mentioned above, has been shown earlier both in device-independent~\cite{Bancal_ST_15, chen_st_16, kaniewski_st_17} and semi device-independent~\cite{Tavakoli_STPM18, Maity_st_21} settings. But in such standard settings, certification of quantum instruments is impossible \cite{Mohan_STI19, Miklin_SDIM20}.} It is only in the sequential settings where certification of non-projective and unsharp measurements \cite{Miklin_SDIM20, Tavakoli_STM20} (instruments) \cite{Mohan_STI19} can be exhibited. Therefore, in our work we focus on the robust certification of quantum instruments in a SDI way in the next subsection. A thorough robustness of the scheme is shown by bounding the estimation of the sharpness parameter $\eta$, from below and above, considering noise in the state preparation, leading to an improvement over earlier results \cite{Mohan_STI19}.

\subsection{Robust certification of sharpness parameter}

In realistic conditions, all  quantum systems deviate from the ideal target description, and hence, less than optimal success probabilities are observed, which can't pinpoint the desired objects. Robustness analysis is the estimation of those objects for the suboptimal observed values. Therefore, any meaningful self-testing argument must come with a robustness analysis.
 Our above-mentioned self-testing statements in Sec.\ref{selftest} are valid only when the optimal trade-off point is achieved, which is not the case in all practical experimental conditions. Here we specifically consider the sharpness parameter of Barun's measurement, which characterizes his instrument. We show how one can certify the sharpness parameter $\eta$, of Barun's measurement in the presence of noise, in a more robust way than the previous work~\cite{Mohan_STI19}. We provide upper and lower bounds on the sharpness parameter for less than optimal observed success probabilities, which merge to the same value in noiseless conditions.

\begin{figure*}[ht]
    \centering

    \begin{subfigure}{0.48\textwidth}
        \centering
        \includegraphics[width=\linewidth]{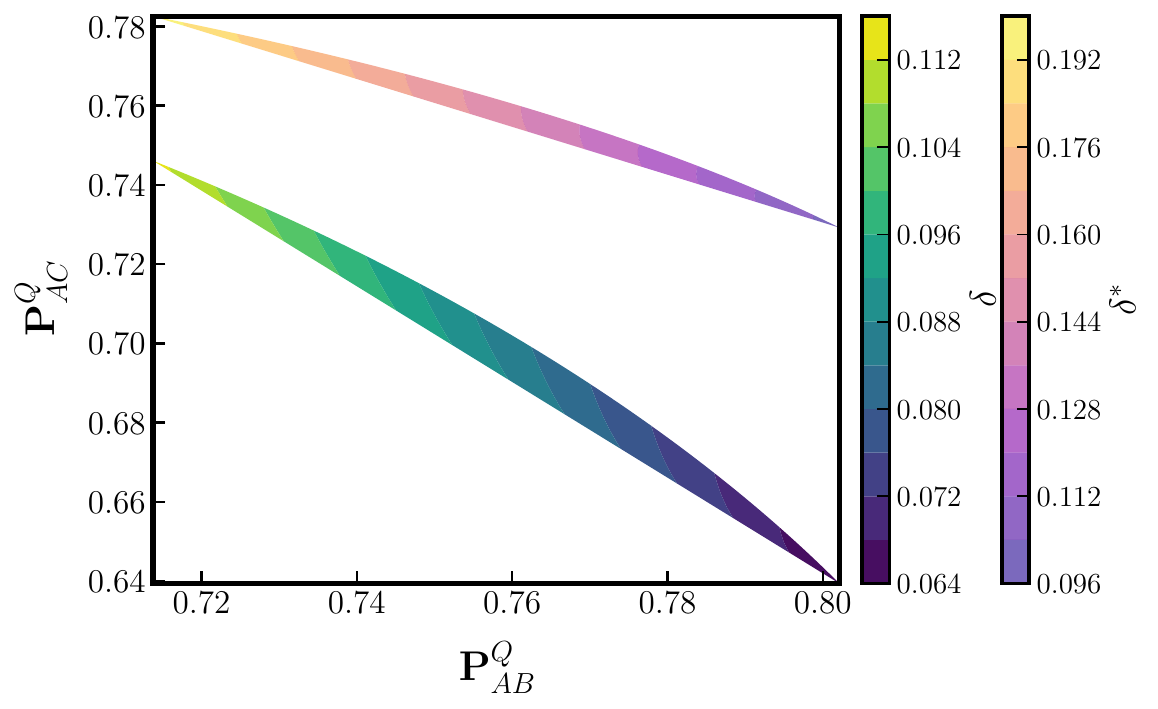}
        \caption{}
        \label{Plot1}
    \end{subfigure}
    \hfill
    \begin{subfigure}{0.48\textwidth}
        \centering
        \includegraphics[width=\linewidth]{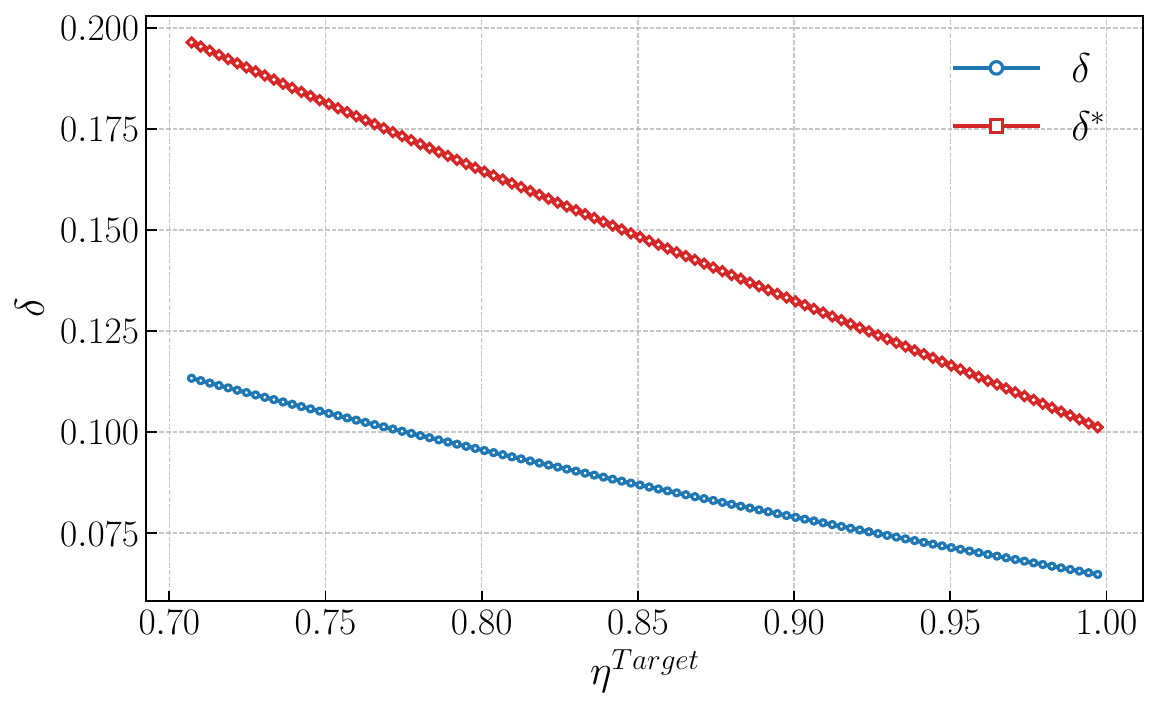}
        \caption{}
        \label{Plot1.1}
    \end{subfigure}
    \caption{
        (a) Contour plots of $\delta$ and $\delta^*$ as functions of $\text{P}_{\text{AB}}^Q$ and $\text{P}_{\text{AC}}^Q$.  Here, $\delta$ denotes the gap between the upper and lower bounds of the sharpness parameter estimated from the two-receiver communication game proposed in this work. On the other hand, $\delta^*$ is the bound gap drawn from the protocol
        without collaboration~\cite{Mohan_STI19}. The noise configurations considered are :  $p_1 = 0.95$, $p_2 = 0.93$, $p_3 = 0.95$; (b) Plots of $\delta$ and $\delta^*$ as functions of $\eta^{Target}$ denoting the value of the sharpness parameter  to be self-tested.  See also Table~\ref{noise_comparison} and Appendix~\ref{ule}.
    }
    \label{Comp_graphs}
\end{figure*}

Any binary outcome qubit observable can be expressed in the following way: $M_y=g_{y0}\mathbb{I}+\vec{g_y}.\vec{\sigma}$. Here, the sharpness parameter of Barun's measurement is equivalent to the length of the Bloch vector $\vec{g_y}$.
For simplicity, we take the  sharpness parameters of Barun's two settings
to be equal, {\it viz.}, $\eta \equiv |\vec{g}_{0}|=|\vec{g}_{1}|$. With these specifications, we can now place a lower bound on $\eta$, based
on the consideration that the success probability of Barun  $P_{AB}^Q$, has to be non-classical. 

In order to bound the sharpness of Barun’s instrument, consider 
Barun's success probability, which  using Eq.~(\ref{PABQB}), is given by,
\begin{equation}
    \begin{aligned}
        P_{AB}^{Q} = \frac{1}{8}(4+|\vec{g}_{0}||\vec{q}_{0}|\hat{q}_{0}\cdot\hat{g}_{0}+|\vec{g}_{1}||\vec{q}_{1}|\hat{q}_{1}\cdot\hat{g}_{1})
    \end{aligned}
\end{equation}
For simplicity, we can now fix the sharpness parameter of Barun's two settings as, $\eta \equiv |\vec{g}_{0}|=|\vec{g}_{1}|$. It follows
that,
\begin{equation}
    \begin{aligned}
        \eta = \frac{8 P_{AB}^{Q} - 4}{|\vec{g}_{0}|\hat{q}_{0}\cdot\hat{g}_{0}+|\vec{g}_{1}|\hat{q}_{1}\cdot\hat{g}_{1}}
    \end{aligned}
\end{equation}
The lower bound of $\eta$ is obtained by maximizing the denominator. This occurs when  $\hat{q}_{0}\cdot\hat{g}_{0}=\hat{q}_{1}\cdot\hat{g}_{1}=1$ and $|\hat{q}_{0}|=|\hat{q}_{1}|$, and one gets
\begin{equation}
    \begin{aligned}
        \eta \geq \sqrt{2}(2 P_{AB}^{Q} -1)
    \end{aligned}
\end{equation}
This bound is non-trivial whenever $P_{AB}^{Q}> 1/2$.

On the other hand, from the requirement that $P_{AC}^Q$ is non-classical, we can place an upper bound on $\eta$.
For obtaining the upper bound of $\eta$, we make use of the optimal choice
of the success probability from Eq.(\ref{tradeoff}), given
\begin{equation}
    \begin{aligned}
        P_{AC}^{Q} \leq \frac{1}{2} + \frac{\sqrt{1-\eta^{2}}}{2\sqrt{2}}
    \end{aligned}
\end{equation}
Inverting the above inequality, leads to,
\begin{equation}
    \begin{aligned}
        \eta \leq \sqrt{8 P_{AC}^Q - 8 (P_{AC}^{Q})^2 - 1}
    \end{aligned}
\end{equation}


\begin{table}[h]
\centering
\begin{tabular}{|l|c|c|c|}
\toprule
\multicolumn{1}{|c|}{\begin{tabular}[c]{@{}c@{}}\textbf{Visibility} \\ \textbf{$(p_1, p_2, p_3)$}\end{tabular}} 
& \begin{tabular}[c]{@{}c@{}}\textbf{Observed} \\ \textbf{Probabilities}\end{tabular} 
& \begin{tabular}[c]{@{}c@{}}$\delta$ \\ $= \eta_{u} - \eta_l$\end{tabular} 
& \begin{tabular}[c]{@{}c@{}}$\delta^*$ \\ $= \eta_{u}^* - \eta_l^*$\end{tabular} \\
\midrule
(0.95, 0.9, 0.95) 
& \begin{tabular}[c]{@{}l@{}}$P_{AB}^{Q} = 0.7138$\\ $P_{AC}^{*Q} = 0.7826$\\ $P_{AC}^{Q} = 0.7461$\\ \end{tabular}
& 0.1133 
& 0.1964 \\
\midrule
(0.98, 0.95, 0.98) 
& \begin{tabular}[c]{@{}l@{}}$P_{AB}^{Q} = 0.7328$\\ $P_{AC}^{*Q} = 0.7955$\\ $P_{AC}^{Q} = 0.7515$\\ \end{tabular}
& 0.0444 
& 0.0824 \\
\midrule
(0.93, 0.88, 0.93) 
& \begin{tabular}[c]{@{}l@{}}$P_{AB}^{Q} = 0.7046$\\ $P_{AC}^{*Q} = 0.7726$\\ $P_{AC}^{Q} = 0.7394$\\ \end{tabular}
& 0.1572 
& 0.2617 \\

\bottomrule
\end{tabular}
\caption{Comparison of estimation-error bands resulting from two different methods (with or without collaboration among the receivers) of self-testing the sharpness parameter. From Fig.\ref{Comp_graphs}, we pick two different sets of values of noise in state preparation and measurements, and the obtained success probabilities of the corresponding games (second column). $P_{AB}^Q$ is the same in both protocols, as the task between the sender and the first receiver is the same as in the standard QRAC. $P_{AC}^{*Q}$ denotes the success probability of the second receiver,
as obtained through the game without collaboration~\cite{Mohan_STI19}, whereas $P_{AC}^{Q}$ indicates our result. In the third column, $\delta$, the gap between the upper and lower bounds on the sharpness parameter resulting
from our result is indicated, and the fourth column refers to the result obtained from
the corresponding analysis in  Ref.~\cite{Mohan_STI19}. The greater robustness of self-testing of the sharpness parameter with respect to our proposed game  
with restricted collaboration between the receivers, compared to the standard sequential QRAC, is clearly revealed.}
\label{noise_comparison}
\end{table}

 From the observed value of $(P_{AB}^Q, P_{AC}^Q)$, the pair, one can estimate the sharpness parameter, which lies within the upper and lower bounds. The narrower the gap between the two bounds, the better the certification. Specifically, we consider a noise model where Aparna's preparation is given by $\rho_{x}^{'} = p_1\rho_{x} + (1 - p_1)\frac{\mathbb{I}}{2}$, and the POVM elements corresponding to the measurement devices of Barun and Chhanda are represented as $M_{y}^{'} = p_2 M_{y} + (1 - p_2)\frac{\mathbb{I}}{2}$ and $N_{z}^{'} = p_3 N_{z} + (1 - p_3)\frac{\mathbb{I}}{2}$, respectively. Here, $p_1$ denotes the visibility of Aparna's preparation device, $p_2$  and  $p_3$  are the sharpness parameter of Barun's and Chhanda’s measurement devices, respectively. In Fig.~\ref{Comp_graphs}, we graphically illustrate the width of the plausible range for the sharpness parameter associated with the underlying noisy measurements. 
 
 In a noisy and suboptimal scenario, the sharpness parameter can be estimated within a significantly narrower range using our proposed communication game compared to the estimation based on the sequential QRAC protocol ~\cite{Mohan_STI19}.
  For instance, suppose Aparna’s preparations have a visibility of 95\%, while Barun’s and Chhanda’s measurements achieve visibilities of 93\% and 95\%, respectively. Under these conditions, our protocol estimates the sharpness parameter with an error gap of approximately $0.1133$, whereas the method in Ref.~\cite{Mohan_STI19} leads to a larger error gap of $0.1964$.
  In Fig.~\ref{Plot1.1}, for the same visibility values, we plot the gap $(\delta, \delta^*)$ as a function of $\eta^{\text{Target}}$, where $\eta^{\text{Target}}$ denotes the value of the sharpness parameter that an experimentalist may want to self-test. Detailed calculations and comparison plots for two other noisy scenarios are provided in Fig.~\ref{Comp_graphs_APP} of  Appendix~\ref{ule}.

In Table~\ref{noise_comparison}, we present a comparison of the gap between the upper and lower bounds, estimated from the observed figures of merit in our proposed communication game and the protocol without such collaboration~\cite{Mohan_STI19}, for three different sets of noise levels. The gap between the bounds over the entire range of allowed observed success probabilities for the visibility set $(p_1 = 0.95, p_2 = 0.9, p_3 = 0.95)$ for the first set of noise levels is illustrated in Fig.~\ref{Plot1}. The results for the other two noisy scenarios are extracted from Fig.~\ref{Comp_graphs_APP} in Appendix~\ref{ule}. It can be seen that the protocol for self-testing the sharpness parameter based on our scheme is more robust compared to the one in Ref.~\cite{Mohan_STI19}
for a range of parameter values.

\section{Sequential communication game in higher dimensions}

\begin{figure*}[ht]
    \centering

    \begin{subfigure}{0.48\textwidth}
        \centering
        \includegraphics[width=\linewidth]{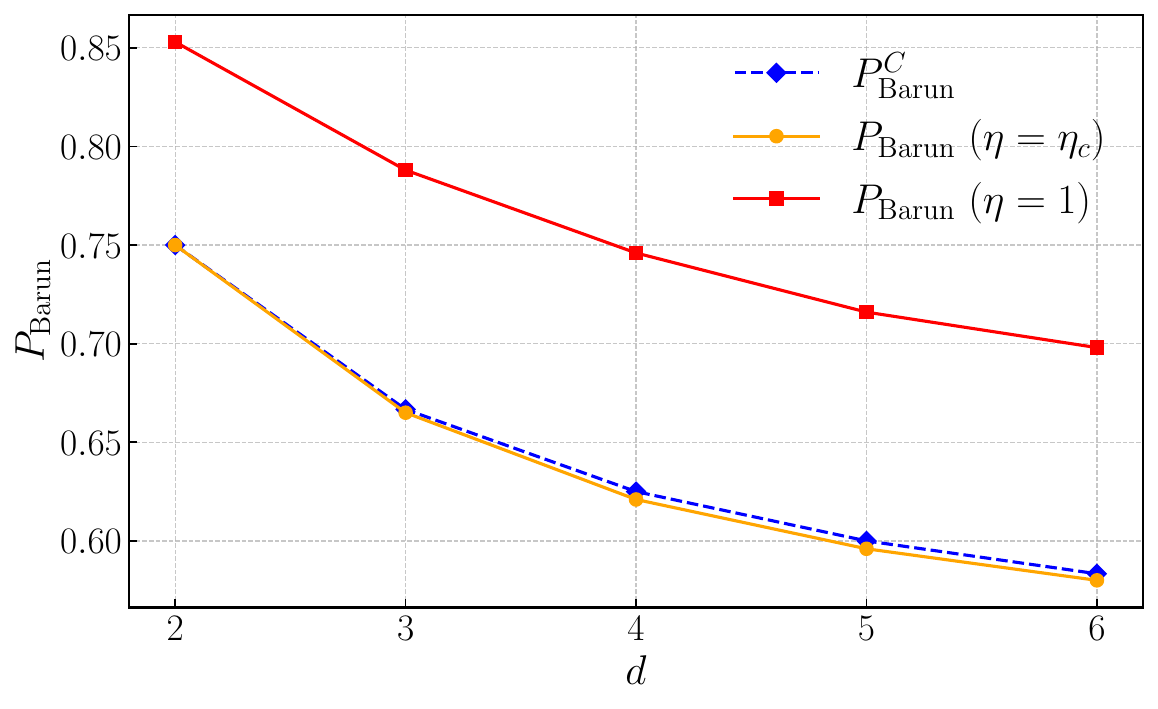}
        \caption{}
        \label{P_Barun_vs_d}
    \end{subfigure}
    \hfill
    \begin{subfigure}{0.48\textwidth}
        \centering
        \includegraphics[width=\linewidth]{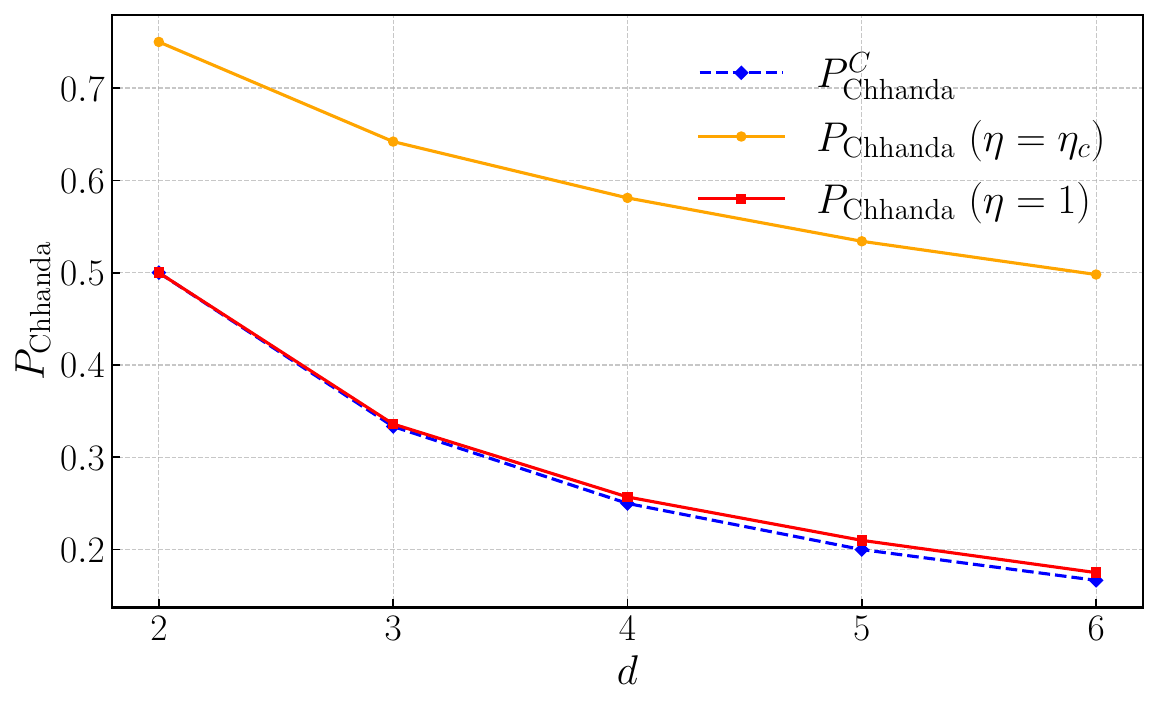}
        \caption{}
        \label{P_Chhanda_vs_d}
    \end{subfigure}

    \vspace{0.5cm}

    \begin{subfigure}{0.48\textwidth}
        \centering
        \includegraphics[width=\linewidth]{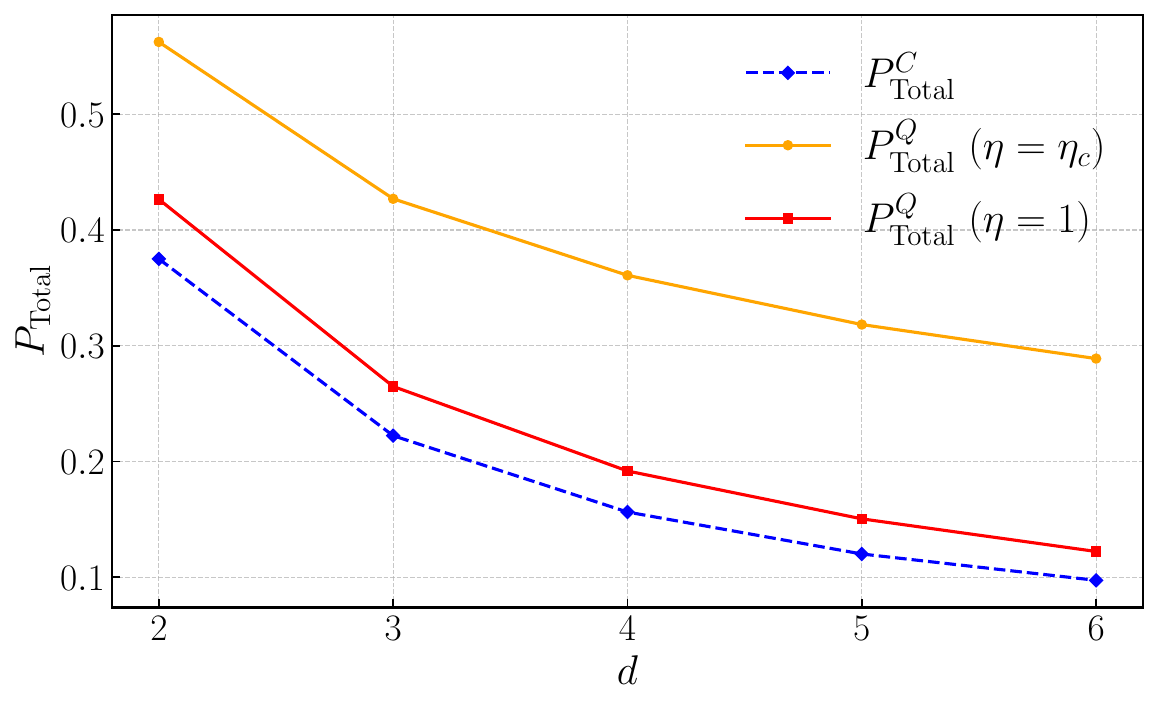}
        \caption{}
        \label{Total_P_vs_D}
    \end{subfigure}

    \caption{ Success probabilities as functions of the system dimension $d$.
        (a) Success probability for Barun.
        (b) Success probability for Channda.
        (c) Total success probability.
        The plots illustrate the quantum advantage of the proposed communication game.
        Notably, quantum strategies utilizing unsharp measurements outperform those restricted to sharp measurements.
        $\eta_c$ corresponds to the critical sharpness parameter.
   }
    \label{Success_Prob_vs_d}
\end{figure*}

Our proposed two-receiver new communication game can be generalized to more sophisticated scenarios of QRAC with higher-dimensional systems and longer messages to be encoded or transmitted \cite{Ambainis_RACSR09, liabotro_ImprovedBnd_17}. Here we discuss the higher-dimensional generalization and compute numerically the lower bound of the optimal success probability employing the see-saw method up to dimension six (see Appendix~\ref{SEE-SAW-Algo} for the details). The see-saw is an iterative optimization method for affine functions of Hermitian operators, which was introduced initially to maximize payoff function of the quantum strategy of a given Bell-type game \cite{wernerwolf_01}. 

The results are displayed in Fig.~\ref{Success_Prob_vs_d} where we separately show the individual payoff functions corresponding Barun and Chhanda and their joint payoff function as described in Section II in the three sub-figures. In each of the sub-figures, we plot the respective success probabilities for three cases: 
(i) the classical success probability $P^C$, (ii) the quantum success probability at the threshold of quantum advantage $P (\eta = \eta_c)$, with 
$\eta_c$ denoting the critical value at which Barun's success probability matches the classical benchmark, given by $\eta_c = \frac{d - 1}{d + \sqrt{d} - 2}$, and (iii) the quantum success probability for sharp measurement by Barun, $P (\eta =1)$. The magnitudes of all the success probabilities are
shown to drop with an increase in dimensions whereas quantum advantage is increasing. Note from 
 Figs.~\ref{P_Barun_vs_d} and~\ref{P_Chhanda_vs_d} that while Barun's success
probability drops with the increase of the sharpness parameter, the reverse
occurs in the case of Chhanda's success probability. Thus, the trade-off between Barun's and Chhanda's success is clearly maintained even in higher dimensions. From Fig.~\ref{Total_P_vs_D} we observe that the magnitude of quantum advantage increases with the dimension. Moreover, the resourcefulness of the unsharp measurement in the context of the present communication task is obvious, outperforming its sharp counterpart.

\section{Conclusions}\label{Salient}

Certification schemes for various quantum components is essential in the context of present-day quantum science and technology, as malicious providers of devices may be a threat to the security and proper functioning of quantum protocols. For certification of quantum states and measurements, there are several schemes in the literature. for example, device-independent self-testing \cite{POPESCU1992411, self-testing_MY_04, Bancal_ST_15, chen_st_16, kaniewski_st_17, Supic2020selftestingof, mal_quantboundary_23}, semi device-independent certification \cite{vsupic2016self, bowles2014one, quintino2015inequivalence, wollmann2016observation, goswami2018one, bian2020experimental} and device dependent protocols \cite{Horodecki_RMP09, GUHNE2009}. It is to be noted that there exist quantum components other than states and measurements which are not certified by the standard above mentioned certification schemes. Recently, sequential extension of the standard protocols has been proposed which can be employed to self-test quantum instruments \cite{Mohan_STI19},  non-projective \cite{Tavakoli_STM20} and unsharp measurements \cite{Miklin_SDIM20} in a semi-device-independent manner going beyond the standard settings of certification of quantum states and measurements.

It may be worth reemphasizing that quantum random access codes are fundamental communication tasks, which can exhibit quantum advantage even for the simplest quantum system — for a qubit. Such advantage is one of the central issues in quantum information science, facilitating many quantum information processing tasks \cite{Pawlowski_SDI11,Brunner_DW13, Tavakoli_CEPM21, Wehner_DW08,Li_SDRNG11, Li_SDRNG12,qBridge_exp_14,saha_incompatible_23}, and is successful in addressing certain fundamental questions as well \cite{spekkens_pc_09, Pawlowski_IC_09, bera_QRAC_22, Mal_FURPC_21, gupta2023quantum}. Non-trivial extensions of the standard QRAC enable self-testing of quantum instruments and non-projective measurements in the presence of noise in realistic experimental scenarios \cite{foletto_ExpSQRAC_20, anwer_expSQRAC_20, xiao_expSQRAC_21,Tavakoli_STM20}.


In the present work, we have proposed a new communication game extending the standard random access code paradigm to the sequential framework \cite{silva_SequentialBell_15, mal_SequentialBell_16, mal_sequentialSteering_18, mal_sequentialEntaglement_18, das2019facets, Akshata_19, Asmita_19, brown2020arbitrarily, cheng2021limitations, mal_SequentialCGLMP_24,
munshi2025device, datta2018sharing, maity2020detection, gupta2021genuine}, involving one sender and two receivers with restricted communication allowed between the receivers. The novel three-party communication game is based on $2\rightarrow 1$ QRAC, which is fundamentally different from the earlier extensions. Our protocol is based on sequential measurements in the prepare-transform-measure scenario,  which has received much attention in recent times \cite{Tavakoli_STPM18, Mironowicz_SDIM19, Maity_st_21, das2022robust,Mohan_STI19, Miklin_SDIM20, debarshi_seq_RAC24}.
We have shown prominent quantum advantage of the game and used it for the purpose of certification of relevant quantum components in a semi device-independent way and generalizes the game to the higher dimensional system as well.

To summarize, in our proposed game, Aparna, the sender, encodes a two-bit message into a bit, and the task of the first receiver, Barun, is to decode a bit value according to his random input, whereas the second receiver, Chhanda, has to work out the leftover message. Thus, Barun and Chhanda may have some non-zero probability of correctly retrieving the full message, and they are allowed to have limited communication: Barun passes his post-measured state without leaking any output information.  
We have found a quantum advantage, which can certify Aparna's state preparation, Barun's instruments, as well as Chhanda's measurement. Remarkably, the SDI certification of the sharpness parameter associated with Barun's instrument is more robust in our protocol compared to that in the method based on sequential QRAC without any collaboration among the receivers~\cite{Mohan_STI19}. 
We further generalize our sequential protocol to higher dimensions. By computing  numerically employing seesaw algorithm, the lower bound on the quantum success probability, we have shown that the quantum advantage increases with system dimensions, highlighting again the role of unsharp measurement in communication tasks relevant for sequential networks.

The analysis presented here gives rise to prospects for future work in several possible directions. It would be interesting to see how robust the certification task based on the proposed quantum access code under restricted communication is in higher dimensions. Further, it would be worthwhile investigating
implications of our proposed game in other RAC-based tasks of foundational and information processing relevance \cite{Pawlowski_SDI11, Brunner_DW13, Tavakoli_CEPM21, Wehner_DW08, Li_SDRNG11, Li_SDRNG12, saha_incompatible_23,     spekkens_pc_09, Pawlowski_IC_09, bera_QRAC_22, Mal_FURPC_21, gupta2023quantum}. Moreover, our present analysis should motivate explorations of more sophisticated generalizations of
the RAC paradigm in the sequential measurement framework, such as in multipartite scenarios involving more than two receivers, towards the goal of harnessing even greater quantum advantage with minimal resources. Finally, one may explore situations in which non-cooperation on the part of the first receiver is allowed by introducing cheating strategies  modeled by adding noise or deforming the post-measurement state  leading to reduction of the payoff functions.
  



\begin{thebibliography}{76}%
	\makeatletter
	\providecommand \@ifxundefined [1]{%
		\@ifx{#1\undefined}
	}%
	\providecommand \@ifnum [1]{%
		\ifnum #1\expandafter \@firstoftwo
		\else \expandafter \@secondoftwo
		\fi
	}%
	\providecommand \@ifx [1]{%
		\ifx #1\expandafter \@firstoftwo
		\else \expandafter \@secondoftwo
		\fi
	}%
	\providecommand \natexlab [1]{#1}%
	\providecommand \enquote  [1]{``#1''}%
	\providecommand \bibnamefont  [1]{#1}%
	\providecommand \bibfnamefont [1]{#1}%
	\providecommand \citenamefont [1]{#1}%
	\providecommand \href@noop [0]{\@secondoftwo}%
	\providecommand \href [0]{\begingroup \@sanitize@url \@href}%
	\providecommand \@href[1]{\@@startlink{#1}\@@href}%
	\providecommand \@@href[1]{\endgroup#1\@@endlink}%
	\providecommand \@sanitize@url [0]{\catcode `\\12\catcode `\$12\catcode
		`\&12\catcode `\#12\catcode `\^12\catcode `\_12\catcode `\%12\relax}%
	\providecommand \@@startlink[1]{}%
	\providecommand \@@endlink[0]{}%
	\providecommand \url  [0]{\begingroup\@sanitize@url \@url }%
	\providecommand \@url [1]{\endgroup\@href {#1}{\urlprefix }}%
	\providecommand \urlprefix  [0]{URL }%
	\providecommand \Eprint [0]{\href }%
	\providecommand \doibase [0]{https://doi.org/}%
	\providecommand \selectlanguage [0]{\@gobble}%
	\providecommand \bibinfo  [0]{\@secondoftwo}%
	\providecommand \bibfield  [0]{\@secondoftwo}%
	\providecommand \translation [1]{[#1]}%
	\providecommand \BibitemOpen [0]{}%
	\providecommand \bibitemStop [0]{}%
	\providecommand \bibitemNoStop [0]{.\EOS\space}%
	\providecommand \EOS [0]{\spacefactor3000\relax}%
	\providecommand \BibitemShut  [1]{\csname bibitem#1\endcsname}%
	\let\auto@bib@innerbib\@empty
	\bibitem [{\citenamefont {Nielsen}\ and\ \citenamefont
		{Chuang}(2010)}]{nielsen_chuang_2010}%
	\BibitemOpen
	\bibfield  {author} {\bibinfo {author} {\bibfnamefont {M.~A.}\ \bibnamefont
			{Nielsen}}\ and\ \bibinfo {author} {\bibfnamefont {I.~L.}\ \bibnamefont
			{Chuang}},\ }\href@noop {} {\emph {\bibinfo {title} {Quantum Computation and
				Quantum Information}}}\ (\bibinfo  {publisher} {Cambridge University Press},\
	\bibinfo {year} {2010})\BibitemShut {NoStop}%
	\bibitem [{\citenamefont {Watrous}(2018)}]{Watruos_QisBook}%
	\BibitemOpen
	\bibfield  {author} {\bibinfo {author} {\bibfnamefont {J.}~\bibnamefont
			{Watrous}},\ }\href@noop {} {\emph {\bibinfo {title} {The Theory of Quantum
				Information}}},\ \bibinfo {edition} {1st}\ ed.\ (\bibinfo  {publisher}
	{Cambridge University Press},\ \bibinfo {address} {USA},\ \bibinfo {year}
	{2018})\BibitemShut {NoStop}%
	\bibitem [{\citenamefont {Hinds}\ and\ \citenamefont {Blatt}(2012)}]{nobel_12}%
	\BibitemOpen
	\bibfield  {author} {\bibinfo {author} {\bibfnamefont {E.}~\bibnamefont
			{Hinds}}\ and\ \bibinfo {author} {\bibfnamefont {R.}~\bibnamefont {Blatt}},\
	}\href {https://doi.org/https://doi.org/10.1038/492055a} {\bibfield
		{journal} {\bibinfo  {journal} {Nature}\ }\textbf {\bibinfo {volume} {492}},\
		\bibinfo {pages} {55} (\bibinfo {year} {2012})}\BibitemShut {NoStop}%
	\bibitem [{\citenamefont {Popescu}\ and\ \citenamefont
		{Rohrlich}(1992)}]{POPESCU1992411}%
	\BibitemOpen
	\bibfield  {author} {\bibinfo {author} {\bibfnamefont {S.}~\bibnamefont
			{Popescu}}\ and\ \bibinfo {author} {\bibfnamefont {D.}~\bibnamefont
			{Rohrlich}},\ }\href
	{https://doi.org/https://doi.org/10.1016/0375-9601(92)90819-8} {\bibfield
		{journal} {\bibinfo  {journal} {Physics Letters A}\ }\textbf {\bibinfo
			{volume} {169}},\ \bibinfo {pages} {411} (\bibinfo {year}
		{1992})}\BibitemShut {NoStop}%
	\bibitem [{\citenamefont {Mayers}\ and\ \citenamefont
		{Yao}(2004)}]{self-testing_MY_04}%
	\BibitemOpen
	\bibfield  {author} {\bibinfo {author} {\bibfnamefont {D.}~\bibnamefont
			{Mayers}}\ and\ \bibinfo {author} {\bibfnamefont {A.}~\bibnamefont {Yao}},\
	}\href {https://dl.acm.org/doi/10.5555/2011827.2011830} {\bibfield  {journal}
		{\bibinfo  {journal} {Quantum Info. Comput.}\ }\textbf {\bibinfo {volume}
			{4}},\ \bibinfo {pages} {273–286} (\bibinfo {year} {2004})}\BibitemShut
	{NoStop}%
	\bibitem [{\citenamefont {Bancal}\ \emph {et~al.}(2015)\citenamefont {Bancal},
		\citenamefont {Navascu\'es}, \citenamefont {Scarani}, \citenamefont
		{V\'ertesi},\ and\ \citenamefont {Yang}}]{Bancal_ST_15}%
	\BibitemOpen
	\bibfield  {author} {\bibinfo {author} {\bibfnamefont {J.-D.}\ \bibnamefont
			{Bancal}}, \bibinfo {author} {\bibfnamefont {M.}~\bibnamefont {Navascu\'es}},
		\bibinfo {author} {\bibfnamefont {V.}~\bibnamefont {Scarani}}, \bibinfo
		{author} {\bibfnamefont {T.}~\bibnamefont {V\'ertesi}},\ and\ \bibinfo
		{author} {\bibfnamefont {T.~H.}\ \bibnamefont {Yang}},\ }\href
	{https://doi.org/10.1103/PhysRevA.91.022115} {\bibfield  {journal} {\bibinfo
			{journal} {Phys. Rev. A}\ }\textbf {\bibinfo {volume} {91}},\ \bibinfo
		{pages} {022115} (\bibinfo {year} {2015})}\BibitemShut {NoStop}%
	\bibitem [{\citenamefont {Chen}\ \emph {et~al.}(2016)\citenamefont {Chen},
		\citenamefont {Budroni}, \citenamefont {Liang},\ and\ \citenamefont
		{Chen}}]{chen_st_16}%
	\BibitemOpen
	\bibfield  {author} {\bibinfo {author} {\bibfnamefont {S.-L.}\ \bibnamefont
			{Chen}}, \bibinfo {author} {\bibfnamefont {C.}~\bibnamefont {Budroni}},
		\bibinfo {author} {\bibfnamefont {Y.-C.}\ \bibnamefont {Liang}},\ and\
		\bibinfo {author} {\bibfnamefont {Y.-N.}\ \bibnamefont {Chen}},\ }\href
	{https://doi.org/10.1103/PhysRevLett.116.240401} {\bibfield  {journal}
		{\bibinfo  {journal} {Phys. Rev. Lett.}\ }\textbf {\bibinfo {volume} {116}},\
		\bibinfo {pages} {240401} (\bibinfo {year} {2016})}\BibitemShut {NoStop}%
	\bibitem [{\citenamefont {Kaniewski}(2017)}]{kaniewski_st_17}%
	\BibitemOpen
	\bibfield  {author} {\bibinfo {author} {\bibfnamefont {J.~m.~k.}\
			\bibnamefont {Kaniewski}},\ }\href
	{https://doi.org/10.1103/PhysRevA.95.062323} {\bibfield  {journal} {\bibinfo
			{journal} {Phys. Rev. A}\ }\textbf {\bibinfo {volume} {95}},\ \bibinfo
		{pages} {062323} (\bibinfo {year} {2017})}\BibitemShut {NoStop}%
	\bibitem [{\citenamefont {{\v{S}}upi{\'{c}}}\ and\ \citenamefont
		{Bowles}(2020)}]{Supic2020selftestingof}%
	\BibitemOpen
	\bibfield  {author} {\bibinfo {author} {\bibfnamefont {I.}~\bibnamefont
			{{\v{S}}upi{\'{c}}}}\ and\ \bibinfo {author} {\bibfnamefont {J.}~\bibnamefont
			{Bowles}},\ }\href {https://doi.org/10.22331/q-2020-09-30-337} {\bibfield
		{journal} {\bibinfo  {journal} {{Quantum}}\ }\textbf {\bibinfo {volume}
			{4}},\ \bibinfo {pages} {337} (\bibinfo {year} {2020})}\BibitemShut {NoStop}%
	\bibitem [{\citenamefont {Chen}\ \emph {et~al.}(2023)\citenamefont {Chen},
		\citenamefont {Tabia}, \citenamefont {Jebarathinam}, \citenamefont {Mal},
		\citenamefont {Wu},\ and\ \citenamefont {Liang}}]{mal_quantboundary_23}%
	\BibitemOpen
	\bibfield  {author} {\bibinfo {author} {\bibfnamefont {K.-S.}\ \bibnamefont
			{Chen}}, \bibinfo {author} {\bibfnamefont {G.~N.~M.}\ \bibnamefont {Tabia}},
		\bibinfo {author} {\bibfnamefont {C.}~\bibnamefont {Jebarathinam}}, \bibinfo
		{author} {\bibfnamefont {S.}~\bibnamefont {Mal}}, \bibinfo {author}
		{\bibfnamefont {J.-Y.}\ \bibnamefont {Wu}},\ and\ \bibinfo {author}
		{\bibfnamefont {Y.-C.}\ \bibnamefont {Liang}},\ }\href
	{https://doi.org/10.22331/q-2023-07-11-1054} {\bibfield  {journal} {\bibinfo
			{journal} {{Quantum}}\ }\textbf {\bibinfo {volume} {7}},\ \bibinfo {pages}
		{1054} (\bibinfo {year} {2023})}\BibitemShut {NoStop}%
	\bibitem [{\citenamefont {{\v{S}}upi{\'c}}\ and\ \citenamefont
		{Hoban}(2016)}]{vsupic2016self}%
	\BibitemOpen
	\bibfield  {author} {\bibinfo {author} {\bibfnamefont {I.}~\bibnamefont
			{{\v{S}}upi{\'c}}}\ and\ \bibinfo {author} {\bibfnamefont {M.~J.}\
			\bibnamefont {Hoban}},\ }\href
	{https://doi.org/10.1088/1367-2630/18/7/075006} {\bibfield  {journal}
		{\bibinfo  {journal} {New Journal of Physics}\ }\textbf {\bibinfo {volume}
			{18}},\ \bibinfo {pages} {075006} (\bibinfo {year} {2016})}\BibitemShut
	{NoStop}%
	\bibitem [{\citenamefont {Bowles}\ \emph {et~al.}(2014)\citenamefont {Bowles},
		\citenamefont {V\'ertesi}, \citenamefont {Quintino},\ and\ \citenamefont
		{Brunner}}]{bowles2014one}%
	\BibitemOpen
	\bibfield  {author} {\bibinfo {author} {\bibfnamefont {J.}~\bibnamefont
			{Bowles}}, \bibinfo {author} {\bibfnamefont {T.}~\bibnamefont {V\'ertesi}},
		\bibinfo {author} {\bibfnamefont {M.~T.}\ \bibnamefont {Quintino}},\ and\
		\bibinfo {author} {\bibfnamefont {N.}~\bibnamefont {Brunner}},\ }\href
	{https://doi.org/10.1103/PhysRevLett.112.200402} {\bibfield  {journal}
		{\bibinfo  {journal} {Phys. Rev. Lett.}\ }\textbf {\bibinfo {volume} {112}},\
		\bibinfo {pages} {200402} (\bibinfo {year} {2014})}\BibitemShut {NoStop}%
	\bibitem [{\citenamefont {Quintino}\ \emph {et~al.}(2015)\citenamefont
		{Quintino}, \citenamefont {V\'ertesi}, \citenamefont {Cavalcanti},
		\citenamefont {Augusiak}, \citenamefont {Demianowicz}, \citenamefont
		{Ac\'{\i}n},\ and\ \citenamefont {Brunner}}]{quintino2015inequivalence}%
	\BibitemOpen
	\bibfield  {author} {\bibinfo {author} {\bibfnamefont {M.~T.}\ \bibnamefont
			{Quintino}}, \bibinfo {author} {\bibfnamefont {T.}~\bibnamefont {V\'ertesi}},
		\bibinfo {author} {\bibfnamefont {D.}~\bibnamefont {Cavalcanti}}, \bibinfo
		{author} {\bibfnamefont {R.}~\bibnamefont {Augusiak}}, \bibinfo {author}
		{\bibfnamefont {M.}~\bibnamefont {Demianowicz}}, \bibinfo {author}
		{\bibfnamefont {A.}~\bibnamefont {Ac\'{\i}n}},\ and\ \bibinfo {author}
		{\bibfnamefont {N.}~\bibnamefont {Brunner}},\ }\href
	{https://doi.org/10.1103/PhysRevA.92.032107} {\bibfield  {journal} {\bibinfo
			{journal} {Phys. Rev. A}\ }\textbf {\bibinfo {volume} {92}},\ \bibinfo
		{pages} {032107} (\bibinfo {year} {2015})}\BibitemShut {NoStop}%
	\bibitem [{\citenamefont {Wollmann}\ \emph {et~al.}(2016)\citenamefont
		{Wollmann}, \citenamefont {Walk}, \citenamefont {Bennet}, \citenamefont
		{Wiseman},\ and\ \citenamefont {Pryde}}]{wollmann2016observation}%
	\BibitemOpen
	\bibfield  {author} {\bibinfo {author} {\bibfnamefont {S.}~\bibnamefont
			{Wollmann}}, \bibinfo {author} {\bibfnamefont {N.}~\bibnamefont {Walk}},
		\bibinfo {author} {\bibfnamefont {A.~J.}\ \bibnamefont {Bennet}}, \bibinfo
		{author} {\bibfnamefont {H.~M.}\ \bibnamefont {Wiseman}},\ and\ \bibinfo
		{author} {\bibfnamefont {G.~J.}\ \bibnamefont {Pryde}},\ }\href
	{https://doi.org/10.1103/PhysRevLett.116.160403} {\bibfield  {journal}
		{\bibinfo  {journal} {Phys. Rev. Lett.}\ }\textbf {\bibinfo {volume} {116}},\
		\bibinfo {pages} {160403} (\bibinfo {year} {2016})}\BibitemShut {NoStop}%
	\bibitem [{\citenamefont {Goswami}\ \emph {et~al.}(2018)\citenamefont
		{Goswami}, \citenamefont {Bhattacharya}, \citenamefont {Das}, \citenamefont
		{Sasmal}, \citenamefont {Jebaratnam},\ and\ \citenamefont
		{Majumdar}}]{goswami2018one}%
	\BibitemOpen
	\bibfield  {author} {\bibinfo {author} {\bibfnamefont {S.}~\bibnamefont
			{Goswami}}, \bibinfo {author} {\bibfnamefont {B.}~\bibnamefont
			{Bhattacharya}}, \bibinfo {author} {\bibfnamefont {D.}~\bibnamefont {Das}},
		\bibinfo {author} {\bibfnamefont {S.}~\bibnamefont {Sasmal}}, \bibinfo
		{author} {\bibfnamefont {C.}~\bibnamefont {Jebaratnam}},\ and\ \bibinfo
		{author} {\bibfnamefont {A.~S.}\ \bibnamefont {Majumdar}},\ }\href
	{https://doi.org/10.1103/PhysRevA.98.022311} {\bibfield  {journal} {\bibinfo
			{journal} {Phys. Rev. A}\ }\textbf {\bibinfo {volume} {98}},\ \bibinfo
		{pages} {022311} (\bibinfo {year} {2018})}\BibitemShut {NoStop}%
	\bibitem [{\citenamefont {Bian}\ \emph {et~al.}(2020)\citenamefont {Bian},
		\citenamefont {Majumdar}, \citenamefont {Jebarathinam}, \citenamefont {Wang},
		\citenamefont {Xiao}, \citenamefont {Zhan}, \citenamefont {Zhang},\ and\
		\citenamefont {Xue}}]{bian2020experimental}%
	\BibitemOpen
	\bibfield  {author} {\bibinfo {author} {\bibfnamefont {Z.}~\bibnamefont
			{Bian}}, \bibinfo {author} {\bibfnamefont {A.~S.}\ \bibnamefont {Majumdar}},
		\bibinfo {author} {\bibfnamefont {C.}~\bibnamefont {Jebarathinam}}, \bibinfo
		{author} {\bibfnamefont {K.}~\bibnamefont {Wang}}, \bibinfo {author}
		{\bibfnamefont {L.}~\bibnamefont {Xiao}}, \bibinfo {author} {\bibfnamefont
			{X.}~\bibnamefont {Zhan}}, \bibinfo {author} {\bibfnamefont {Y.}~\bibnamefont
			{Zhang}},\ and\ \bibinfo {author} {\bibfnamefont {P.}~\bibnamefont {Xue}},\
	}\href {https://doi.org/10.1103/PhysRevA.101.020301} {\bibfield  {journal}
		{\bibinfo  {journal} {Phys. Rev. A}\ }\textbf {\bibinfo {volume} {101}},\
		\bibinfo {pages} {020301} (\bibinfo {year} {2020})}\BibitemShut {NoStop}%
	\bibitem [{\citenamefont {Horodecki}\ \emph {et~al.}(2009)\citenamefont
		{Horodecki}, \citenamefont {Horodecki}, \citenamefont {Horodecki},\ and\
		\citenamefont {Horodecki}}]{Horodecki_RMP09}%
	\BibitemOpen
	\bibfield  {author} {\bibinfo {author} {\bibfnamefont {R.}~\bibnamefont
			{Horodecki}}, \bibinfo {author} {\bibfnamefont {P.}~\bibnamefont
			{Horodecki}}, \bibinfo {author} {\bibfnamefont {M.}~\bibnamefont
			{Horodecki}},\ and\ \bibinfo {author} {\bibfnamefont {K.}~\bibnamefont
			{Horodecki}},\ }\href {https://doi.org/10.1103/RevModPhys.81.865} {\bibfield
		{journal} {\bibinfo  {journal} {Rev. Mod. Phys.}\ }\textbf {\bibinfo {volume}
			{81}},\ \bibinfo {pages} {865} (\bibinfo {year} {2009})}\BibitemShut
	{NoStop}%
	\bibitem [{\citenamefont {Gühne}\ and\ \citenamefont
		{Tóth}(2009)}]{GUHNE2009}%
	\BibitemOpen
	\bibfield  {author} {\bibinfo {author} {\bibfnamefont {O.}~\bibnamefont
			{Gühne}}\ and\ \bibinfo {author} {\bibfnamefont {G.}~\bibnamefont {Tóth}},\
	}\href {https://doi.org/https://doi.org/10.1016/j.physrep.2009.02.004}
	{\bibfield  {journal} {\bibinfo  {journal} {Physics Reports}\ }\textbf
		{\bibinfo {volume} {474}},\ \bibinfo {pages} {1} (\bibinfo {year}
		{2009})}\BibitemShut {NoStop}%
	\bibitem [{\citenamefont {Holevo}(1973)}]{Holevo73}%
	\BibitemOpen
	\bibfield  {author} {\bibinfo {author} {\bibfnamefont {A.~S.}\ \bibnamefont
			{Holevo}},\ }\href {https://doi.org/https://www.mathnet.ru/eng/ppi903}
	{\bibfield  {journal} {\bibinfo  {journal} {Probl. Peredachi Inf.}\ }\textbf
		{\bibinfo {volume} {9}},\ \bibinfo {pages} {3} (\bibinfo {year}
		{1973})}\BibitemShut {NoStop}%
	\bibitem [{\citenamefont {Wiesner}(1983)}]{wisener_83}%
	\BibitemOpen
	\bibfield  {author} {\bibinfo {author} {\bibfnamefont {S.}~\bibnamefont
			{Wiesner}},\ }\href {https://doi.org/10.1145/1008908.1008920} {\bibfield
		{journal} {\bibinfo  {journal} {SIGACT News}\ }\textbf {\bibinfo {volume}
			{15}},\ \bibinfo {pages} {78–88} (\bibinfo {year} {1983})}\BibitemShut
	{NoStop}%
	\bibitem [{\citenamefont {Ambainis}\ \emph {et~al.}(1999)\citenamefont
		{Ambainis}, \citenamefont {Nayak}, \citenamefont {Ta-Shma},\ and\
		\citenamefont {Vazirani}}]{Ambainis_DCQA99}%
	\BibitemOpen
	\bibfield  {author} {\bibinfo {author} {\bibfnamefont {A.}~\bibnamefont
			{Ambainis}}, \bibinfo {author} {\bibfnamefont {A.}~\bibnamefont {Nayak}},
		\bibinfo {author} {\bibfnamefont {A.}~\bibnamefont {Ta-Shma}},\ and\ \bibinfo
		{author} {\bibfnamefont {U.}~\bibnamefont {Vazirani}},\ }in\ \href
	{https://doi.org/10.1145/301250.301347} {\emph {\bibinfo {booktitle}
			{Proceedings of the Thirty-First Annual ACM Symposium on Theory of
				Computing}}},\ \bibinfo {series and number} {STOC '99}\ (\bibinfo
	{publisher} {Association for Computing Machinery},\ \bibinfo {address} {New
		York, NY, USA},\ \bibinfo {year} {1999})\ p.\ \bibinfo {pages}
	{376–383}\BibitemShut {NoStop}%
	\bibitem [{\citenamefont {Ambainis}\ \emph {et~al.}(2002)\citenamefont
		{Ambainis}, \citenamefont {Nayak}, \citenamefont {Ta-Shma},\ and\
		\citenamefont {Vazirani}}]{Ambainis_DCFA02}%
	\BibitemOpen
	\bibfield  {author} {\bibinfo {author} {\bibfnamefont {A.}~\bibnamefont
			{Ambainis}}, \bibinfo {author} {\bibfnamefont {A.}~\bibnamefont {Nayak}},
		\bibinfo {author} {\bibfnamefont {A.}~\bibnamefont {Ta-Shma}},\ and\ \bibinfo
		{author} {\bibfnamefont {U.}~\bibnamefont {Vazirani}},\ }\href
	{https://doi.org/10.1145/581771.581773} {\bibfield  {journal} {\bibinfo
			{journal} {J. ACM}\ }\textbf {\bibinfo {volume} {49}},\ \bibinfo {pages}
		{496–511} (\bibinfo {year} {2002})}\BibitemShut {NoStop}%
	\bibitem [{\citenamefont {Ambainis}\ \emph {et~al.}(2009)\citenamefont
		{Ambainis}, \citenamefont {Leung}, \citenamefont {Mancinska},\ and\
		\citenamefont {Ozols}}]{Ambainis_RACSR09}%
	\BibitemOpen
	\bibfield  {author} {\bibinfo {author} {\bibfnamefont {A.}~\bibnamefont
			{Ambainis}}, \bibinfo {author} {\bibfnamefont {D.}~\bibnamefont {Leung}},
		\bibinfo {author} {\bibfnamefont {L.}~\bibnamefont {Mancinska}},\ and\
		\bibinfo {author} {\bibfnamefont {M.}~\bibnamefont {Ozols}},\ }\href
	{https://arxiv.org/abs/0810.2937} {\bibinfo {title} {Quantum random access
			codes with shared randomness}} (\bibinfo {year} {2009}),\ \Eprint
	{https://arxiv.org/abs/0810.2937} {arXiv:0810.2937 [quant-ph]} \BibitemShut
	{NoStop}%
	\bibitem [{\citenamefont {Tavakoli}\ \emph {et~al.}(2015)\citenamefont
		{Tavakoli}, \citenamefont {Hameedi}, \citenamefont {Marques},\ and\
		\citenamefont {Bourennane}}]{Tavakoli_QRAC15}%
	\BibitemOpen
	\bibfield  {author} {\bibinfo {author} {\bibfnamefont {A.}~\bibnamefont
			{Tavakoli}}, \bibinfo {author} {\bibfnamefont {A.}~\bibnamefont {Hameedi}},
		\bibinfo {author} {\bibfnamefont {B.}~\bibnamefont {Marques}},\ and\ \bibinfo
		{author} {\bibfnamefont {M.}~\bibnamefont {Bourennane}},\ }\href
	{https://doi.org/10.1103/PhysRevLett.114.170502} {\bibfield  {journal}
		{\bibinfo  {journal} {Phys. Rev. Lett.}\ }\textbf {\bibinfo {volume} {114}},\
		\bibinfo {pages} {170502} (\bibinfo {year} {2015})}\BibitemShut {NoStop}%
	\bibitem [{\citenamefont {Paw\l{}owski}\ and\ \citenamefont
		{Brunner}(2011)}]{Pawlowski_SDI11}%
	\BibitemOpen
	\bibfield  {author} {\bibinfo {author} {\bibfnamefont {M.}~\bibnamefont
			{Paw\l{}owski}}\ and\ \bibinfo {author} {\bibfnamefont {N.}~\bibnamefont
			{Brunner}},\ }\href {https://doi.org/10.1103/PhysRevA.84.010302} {\bibfield
		{journal} {\bibinfo  {journal} {Phys. Rev. A}\ }\textbf {\bibinfo {volume}
			{84}},\ \bibinfo {pages} {010302} (\bibinfo {year} {2011})}\BibitemShut
	{NoStop}%
	\bibitem [{\citenamefont {Brunner}\ \emph {et~al.}(2013)\citenamefont
		{Brunner}, \citenamefont {Navascu\'es},\ and\ \citenamefont
		{V\'ertesi}}]{Brunner_DW13}%
	\BibitemOpen
	\bibfield  {author} {\bibinfo {author} {\bibfnamefont {N.}~\bibnamefont
			{Brunner}}, \bibinfo {author} {\bibfnamefont {M.}~\bibnamefont
			{Navascu\'es}},\ and\ \bibinfo {author} {\bibfnamefont {T.}~\bibnamefont
			{V\'ertesi}},\ }\href {https://doi.org/10.1103/PhysRevLett.110.150501}
	{\bibfield  {journal} {\bibinfo  {journal} {Phys. Rev. Lett.}\ }\textbf
		{\bibinfo {volume} {110}},\ \bibinfo {pages} {150501} (\bibinfo {year}
		{2013})}\BibitemShut {NoStop}%
	\bibitem [{\citenamefont {Tavakoli}\ \emph {et~al.}(2021)\citenamefont
		{Tavakoli}, \citenamefont {Pauwels}, \citenamefont {Woodhead},\ and\
		\citenamefont {Pironio}}]{Tavakoli_CEPM21}%
	\BibitemOpen
	\bibfield  {author} {\bibinfo {author} {\bibfnamefont {A.}~\bibnamefont
			{Tavakoli}}, \bibinfo {author} {\bibfnamefont {J.}~\bibnamefont {Pauwels}},
		\bibinfo {author} {\bibfnamefont {E.}~\bibnamefont {Woodhead}},\ and\
		\bibinfo {author} {\bibfnamefont {S.}~\bibnamefont {Pironio}},\ }\href
	{https://doi.org/10.1103/PRXQuantum.2.040357} {\bibfield  {journal} {\bibinfo
			{journal} {PRX Quantum}\ }\textbf {\bibinfo {volume} {2}},\ \bibinfo {pages}
		{040357} (\bibinfo {year} {2021})}\BibitemShut {NoStop}%
	\bibitem [{\citenamefont {Wehner}\ \emph {et~al.}(2008)\citenamefont {Wehner},
		\citenamefont {Christandl},\ and\ \citenamefont {Doherty}}]{Wehner_DW08}%
	\BibitemOpen
	\bibfield  {author} {\bibinfo {author} {\bibfnamefont {S.}~\bibnamefont
			{Wehner}}, \bibinfo {author} {\bibfnamefont {M.}~\bibnamefont {Christandl}},\
		and\ \bibinfo {author} {\bibfnamefont {A.~C.}\ \bibnamefont {Doherty}},\
	}\href {https://doi.org/10.1103/PhysRevA.78.062112} {\bibfield  {journal}
		{\bibinfo  {journal} {Phys. Rev. A}\ }\textbf {\bibinfo {volume} {78}},\
		\bibinfo {pages} {062112} (\bibinfo {year} {2008})}\BibitemShut {NoStop}%
	\bibitem [{\citenamefont {Li}\ \emph {et~al.}(2011)\citenamefont {Li},
		\citenamefont {Yin}, \citenamefont {Wu}, \citenamefont {Zou}, \citenamefont
		{Wang}, \citenamefont {Chen}, \citenamefont {Guo},\ and\ \citenamefont
		{Han}}]{Li_SDRNG11}%
	\BibitemOpen
	\bibfield  {author} {\bibinfo {author} {\bibfnamefont {H.-W.}\ \bibnamefont
			{Li}}, \bibinfo {author} {\bibfnamefont {Z.-Q.}\ \bibnamefont {Yin}},
		\bibinfo {author} {\bibfnamefont {Y.-C.}\ \bibnamefont {Wu}}, \bibinfo
		{author} {\bibfnamefont {X.-B.}\ \bibnamefont {Zou}}, \bibinfo {author}
		{\bibfnamefont {S.}~\bibnamefont {Wang}}, \bibinfo {author} {\bibfnamefont
			{W.}~\bibnamefont {Chen}}, \bibinfo {author} {\bibfnamefont {G.-C.}\
			\bibnamefont {Guo}},\ and\ \bibinfo {author} {\bibfnamefont {Z.-F.}\
			\bibnamefont {Han}},\ }\href {https://doi.org/10.1103/PhysRevA.84.034301}
	{\bibfield  {journal} {\bibinfo  {journal} {Phys. Rev. A}\ }\textbf {\bibinfo
			{volume} {84}},\ \bibinfo {pages} {034301} (\bibinfo {year}
		{2011})}\BibitemShut {NoStop}%
	\bibitem [{\citenamefont {Li}\ \emph {et~al.}(2012)\citenamefont {Li},
		\citenamefont {Paw\l{}owski}, \citenamefont {Yin}, \citenamefont {Guo},\ and\
		\citenamefont {Han}}]{Li_SDRNG12}%
	\BibitemOpen
	\bibfield  {author} {\bibinfo {author} {\bibfnamefont {H.-W.}\ \bibnamefont
			{Li}}, \bibinfo {author} {\bibfnamefont {M.}~\bibnamefont {Paw\l{}owski}},
		\bibinfo {author} {\bibfnamefont {Z.-Q.}\ \bibnamefont {Yin}}, \bibinfo
		{author} {\bibfnamefont {G.-C.}\ \bibnamefont {Guo}},\ and\ \bibinfo {author}
		{\bibfnamefont {Z.-F.}\ \bibnamefont {Han}},\ }\href
	{https://doi.org/10.1103/PhysRevA.85.052308} {\bibfield  {journal} {\bibinfo
			{journal} {Phys. Rev. A}\ }\textbf {\bibinfo {volume} {85}},\ \bibinfo
		{pages} {052308} (\bibinfo {year} {2012})}\BibitemShut {NoStop}%
	\bibitem [{\citenamefont {Muhammad}\ \emph {et~al.}(2014)\citenamefont
		{Muhammad}, \citenamefont {Tavakoli}, \citenamefont {Kurant}, \citenamefont
		{Paw\l{}owski}, \citenamefont {\ifmmode~\dot{Z}\else \.{Z}\fi{}ukowski},\
		and\ \citenamefont {Bourennane}}]{qBridge_exp_14}%
	\BibitemOpen
	\bibfield  {author} {\bibinfo {author} {\bibfnamefont {S.}~\bibnamefont
			{Muhammad}}, \bibinfo {author} {\bibfnamefont {A.}~\bibnamefont {Tavakoli}},
		\bibinfo {author} {\bibfnamefont {M.}~\bibnamefont {Kurant}}, \bibinfo
		{author} {\bibfnamefont {M.}~\bibnamefont {Paw\l{}owski}}, \bibinfo {author}
		{\bibfnamefont {M.}~\bibnamefont {\ifmmode~\dot{Z}\else \.{Z}\fi{}ukowski}},\
		and\ \bibinfo {author} {\bibfnamefont {M.}~\bibnamefont {Bourennane}},\
	}\href {https://doi.org/10.1103/PhysRevX.4.021047} {\bibfield  {journal}
		{\bibinfo  {journal} {Phys. Rev. X}\ }\textbf {\bibinfo {volume} {4}},\
		\bibinfo {pages} {021047} (\bibinfo {year} {2014})}\BibitemShut {NoStop}%
	\bibitem [{\citenamefont {Saha}\ \emph {et~al.}(2023)\citenamefont {Saha},
		\citenamefont {Das}, \citenamefont {Das}, \citenamefont {Bhattacharya},\ and\
		\citenamefont {Majumdar}}]{saha_incompatible_23}%
	\BibitemOpen
	\bibfield  {author} {\bibinfo {author} {\bibfnamefont {D.}~\bibnamefont
			{Saha}}, \bibinfo {author} {\bibfnamefont {D.}~\bibnamefont {Das}}, \bibinfo
		{author} {\bibfnamefont {A.~K.}\ \bibnamefont {Das}}, \bibinfo {author}
		{\bibfnamefont {B.}~\bibnamefont {Bhattacharya}},\ and\ \bibinfo {author}
		{\bibfnamefont {A.~S.}\ \bibnamefont {Majumdar}},\ }\href
	{https://doi.org/10.1103/PhysRevA.107.062210} {\bibfield  {journal} {\bibinfo
			{journal} {Phys. Rev. A}\ }\textbf {\bibinfo {volume} {107}},\ \bibinfo
		{pages} {062210} (\bibinfo {year} {2023})}\BibitemShut {NoStop}%
	\bibitem [{\citenamefont {Tavakoli}\ \emph {et~al.}(2018)\citenamefont
		{Tavakoli}, \citenamefont {Kaniewski}, \citenamefont {V\'ertesi},
		\citenamefont {Rosset},\ and\ \citenamefont {Brunner}}]{Tavakoli_STPM18}%
	\BibitemOpen
	\bibfield  {author} {\bibinfo {author} {\bibfnamefont {A.}~\bibnamefont
			{Tavakoli}}, \bibinfo {author} {\bibfnamefont {J.~m.~k.}\ \bibnamefont
			{Kaniewski}}, \bibinfo {author} {\bibfnamefont {T.}~\bibnamefont
			{V\'ertesi}}, \bibinfo {author} {\bibfnamefont {D.}~\bibnamefont {Rosset}},\
		and\ \bibinfo {author} {\bibfnamefont {N.}~\bibnamefont {Brunner}},\ }\href
	{https://doi.org/10.1103/PhysRevA.98.062307} {\bibfield  {journal} {\bibinfo
			{journal} {Phys. Rev. A}\ }\textbf {\bibinfo {volume} {98}},\ \bibinfo
		{pages} {062307} (\bibinfo {year} {2018})}\BibitemShut {NoStop}%
	\bibitem [{\citenamefont {Mironowicz}\ and\ \citenamefont
		{Paw\l{}owski}(2019)}]{Mironowicz_SDIM19}%
	\BibitemOpen
	\bibfield  {author} {\bibinfo {author} {\bibfnamefont {P.}~\bibnamefont
			{Mironowicz}}\ and\ \bibinfo {author} {\bibfnamefont {M.}~\bibnamefont
			{Paw\l{}owski}},\ }\href {https://doi.org/10.1103/PhysRevA.100.030301}
	{\bibfield  {journal} {\bibinfo  {journal} {Phys. Rev. A}\ }\textbf {\bibinfo
			{volume} {100}},\ \bibinfo {pages} {030301} (\bibinfo {year}
		{2019})}\BibitemShut {NoStop}%
	\bibitem [{\citenamefont {Maity}\ \emph {et~al.}(2021)\citenamefont {Maity},
		\citenamefont {Mal}, \citenamefont {Jebarathinam},\ and\ \citenamefont
		{Majumdar}}]{Maity_st_21}%
	\BibitemOpen
	\bibfield  {author} {\bibinfo {author} {\bibfnamefont {A.~G.}\ \bibnamefont
			{Maity}}, \bibinfo {author} {\bibfnamefont {S.}~\bibnamefont {Mal}}, \bibinfo
		{author} {\bibfnamefont {C.}~\bibnamefont {Jebarathinam}},\ and\ \bibinfo
		{author} {\bibfnamefont {A.~S.}\ \bibnamefont {Majumdar}},\ }\href
	{https://doi.org/10.1103/PhysRevA.103.062604} {\bibfield  {journal} {\bibinfo
			{journal} {Phys. Rev. A}\ }\textbf {\bibinfo {volume} {103}},\ \bibinfo
		{pages} {062604} (\bibinfo {year} {2021})}\BibitemShut {NoStop}%
	\bibitem [{\citenamefont {Das}\ \emph {et~al.}(2022{\natexlab{a}})\citenamefont
		{Das}, \citenamefont {Maity}, \citenamefont {Saha},\ and\ \citenamefont
		{Majumdar}}]{das2022robust}%
	\BibitemOpen
	\bibfield  {author} {\bibinfo {author} {\bibfnamefont {D.}~\bibnamefont
			{Das}}, \bibinfo {author} {\bibfnamefont {A.~G.}\ \bibnamefont {Maity}},
		\bibinfo {author} {\bibfnamefont {D.}~\bibnamefont {Saha}},\ and\ \bibinfo
		{author} {\bibfnamefont {A.~S.}\ \bibnamefont {Majumdar}},\ }\href@noop {}
	{\bibfield  {journal} {\bibinfo  {journal} {Quantum}\ }\textbf {\bibinfo
			{volume} {6}},\ \bibinfo {pages} {716} (\bibinfo {year}
		{2022}{\natexlab{a}})}\BibitemShut {NoStop}%
	\bibitem [{\citenamefont {Bell}(1964)}]{Bell_64}%
	\BibitemOpen
	\bibfield  {author} {\bibinfo {author} {\bibfnamefont {J.~S.}\ \bibnamefont
			{Bell}},\ }\href {https://doi.org/10.1103/PhysicsPhysiqueFizika.1.195}
	{\bibfield  {journal} {\bibinfo  {journal} {Physics Physique Fizika}\
		}\textbf {\bibinfo {volume} {1}},\ \bibinfo {pages} {195} (\bibinfo {year}
		{1964})}\BibitemShut {NoStop}%
	\bibitem [{\citenamefont {Clauser}\ \emph {et~al.}(1969)\citenamefont
		{Clauser}, \citenamefont {Horne}, \citenamefont {Shimony},\ and\
		\citenamefont {Holt}}]{chsh_69}%
	\BibitemOpen
	\bibfield  {author} {\bibinfo {author} {\bibfnamefont {J.~F.}\ \bibnamefont
			{Clauser}}, \bibinfo {author} {\bibfnamefont {M.~A.}\ \bibnamefont {Horne}},
		\bibinfo {author} {\bibfnamefont {A.}~\bibnamefont {Shimony}},\ and\ \bibinfo
		{author} {\bibfnamefont {R.~A.}\ \bibnamefont {Holt}},\ }\href
	{https://doi.org/10.1103/PhysRevLett.23.880} {\bibfield  {journal} {\bibinfo
			{journal} {Phys. Rev. Lett.}\ }\textbf {\bibinfo {volume} {23}},\ \bibinfo
		{pages} {880} (\bibinfo {year} {1969})}\BibitemShut {NoStop}%
	\bibitem [{\citenamefont {Spekkens}\ \emph {et~al.}(2009)\citenamefont
		{Spekkens}, \citenamefont {Buzacott}, \citenamefont {Keehn}, \citenamefont
		{Toner},\ and\ \citenamefont {Pryde}}]{spekkens_pc_09}%
	\BibitemOpen
	\bibfield  {author} {\bibinfo {author} {\bibfnamefont {R.~W.}\ \bibnamefont
			{Spekkens}}, \bibinfo {author} {\bibfnamefont {D.~H.}\ \bibnamefont
			{Buzacott}}, \bibinfo {author} {\bibfnamefont {A.~J.}\ \bibnamefont {Keehn}},
		\bibinfo {author} {\bibfnamefont {B.}~\bibnamefont {Toner}},\ and\ \bibinfo
		{author} {\bibfnamefont {G.~J.}\ \bibnamefont {Pryde}},\ }\href
	{https://doi.org/10.1103/PhysRevLett.102.010401} {\bibfield  {journal}
		{\bibinfo  {journal} {Phys. Rev. Lett.}\ }\textbf {\bibinfo {volume} {102}},\
		\bibinfo {pages} {010401} (\bibinfo {year} {2009})}\BibitemShut {NoStop}%
	\bibitem [{\citenamefont {Paw{\l}owski}\ \emph {et~al.}(2009)\citenamefont
		{Paw{\l}owski}, \citenamefont {Paterek}, \citenamefont {Kaszlikowski},
		\citenamefont {Scarani}, \citenamefont {Winter},\ and\ \citenamefont
		{Zukowski}}]{Pawlowski_IC_09}%
	\BibitemOpen
	\bibfield  {author} {\bibinfo {author} {\bibfnamefont {M.}~\bibnamefont
			{Paw{\l}owski}}, \bibinfo {author} {\bibfnamefont {T.}~\bibnamefont
			{Paterek}}, \bibinfo {author} {\bibfnamefont {D.}~\bibnamefont
			{Kaszlikowski}}, \bibinfo {author} {\bibfnamefont {V.}~\bibnamefont
			{Scarani}}, \bibinfo {author} {\bibfnamefont {A.}~\bibnamefont {Winter}},\
		and\ \bibinfo {author} {\bibfnamefont {M.}~\bibnamefont {Zukowski}},\
	}\href@noop {} {\bibfield  {journal} {\bibinfo  {journal} {Nature}\ }\textbf
		{\bibinfo {volume} {461}},\ \bibinfo {pages} {1101} (\bibinfo {year}
		{2009})}\BibitemShut {NoStop}%
	\bibitem [{\citenamefont {Bera}\ \emph {et~al.}(2022)\citenamefont {Bera},
		\citenamefont {Maity}, \citenamefont {Mal},\ and\ \citenamefont
		{Majumdar}}]{bera_QRAC_22}%
	\BibitemOpen
	\bibfield  {author} {\bibinfo {author} {\bibfnamefont {S.}~\bibnamefont
			{Bera}}, \bibinfo {author} {\bibfnamefont {A.~G.}\ \bibnamefont {Maity}},
		\bibinfo {author} {\bibfnamefont {S.}~\bibnamefont {Mal}},\ and\ \bibinfo
		{author} {\bibfnamefont {A.~S.}\ \bibnamefont {Majumdar}},\ }\href
	{https://doi.org/10.1103/PhysRevA.106.042439} {\bibfield  {journal} {\bibinfo
			{journal} {Phys. Rev. A}\ }\textbf {\bibinfo {volume} {106}},\ \bibinfo
		{pages} {042439} (\bibinfo {year} {2022})}\BibitemShut {NoStop}%
	\bibitem [{\citenamefont {Sharma}\ \emph {et~al.}(2021)\citenamefont {Sharma},
		\citenamefont {Sazim},\ and\ \citenamefont {Mal}}]{Mal_FURPC_21}%
	\BibitemOpen
	\bibfield  {author} {\bibinfo {author} {\bibfnamefont {G.}~\bibnamefont
			{Sharma}}, \bibinfo {author} {\bibfnamefont {S.}~\bibnamefont {Sazim}},\ and\
		\bibinfo {author} {\bibfnamefont {S.}~\bibnamefont {Mal}},\ }\href
	{https://doi.org/10.1103/PhysRevA.104.032424} {\bibfield  {journal} {\bibinfo
			{journal} {Phys. Rev. A}\ }\textbf {\bibinfo {volume} {104}},\ \bibinfo
		{pages} {032424} (\bibinfo {year} {2021})}\BibitemShut {NoStop}%
	\bibitem [{\citenamefont {Gupta}\ \emph {et~al.}(2023)\citenamefont {Gupta},
		\citenamefont {Saha}, \citenamefont {Xu}, \citenamefont {Cabello},\ and\
		\citenamefont {Majumdar}}]{gupta2023quantum}%
	\BibitemOpen
	\bibfield  {author} {\bibinfo {author} {\bibfnamefont {S.}~\bibnamefont
			{Gupta}}, \bibinfo {author} {\bibfnamefont {D.}~\bibnamefont {Saha}},
		\bibinfo {author} {\bibfnamefont {Z.-P.}\ \bibnamefont {Xu}}, \bibinfo
		{author} {\bibfnamefont {A.}~\bibnamefont {Cabello}},\ and\ \bibinfo {author}
		{\bibfnamefont {A.~S.}\ \bibnamefont {Majumdar}},\ }\href
	{https://doi.org/10.1103/PhysRevLett.130.080802} {\bibfield  {journal}
		{\bibinfo  {journal} {Phys. Rev. Lett.}\ }\textbf {\bibinfo {volume} {130}},\
		\bibinfo {pages} {080802} (\bibinfo {year} {2023})}\BibitemShut {NoStop}%
	\bibitem [{\citenamefont {Aguilar}\ \emph {et~al.}(2018)\citenamefont
		{Aguilar}, \citenamefont {Borka\l{}a}, \citenamefont {Mironowicz},\ and\
		\citenamefont {Paw\l{}owski}}]{Aguilar_pqrac_18}%
	\BibitemOpen
	\bibfield  {author} {\bibinfo {author} {\bibfnamefont {E.~A.}\ \bibnamefont
			{Aguilar}}, \bibinfo {author} {\bibfnamefont {J.~J.}\ \bibnamefont
			{Borka\l{}a}}, \bibinfo {author} {\bibfnamefont {P.}~\bibnamefont
			{Mironowicz}},\ and\ \bibinfo {author} {\bibfnamefont {M.}~\bibnamefont
			{Paw\l{}owski}},\ }\href {https://doi.org/10.1103/PhysRevLett.121.050501}
	{\bibfield  {journal} {\bibinfo  {journal} {Phys. Rev. Lett.}\ }\textbf
		{\bibinfo {volume} {121}},\ \bibinfo {pages} {050501} (\bibinfo {year}
		{2018})}\BibitemShut {NoStop}%
	\bibitem [{\citenamefont {Foletto}\ \emph {et~al.}(2020)\citenamefont
		{Foletto}, \citenamefont {Calderaro}, \citenamefont {Vallone},\ and\
		\citenamefont {Villoresi}}]{foletto_ExpSQRAC_20}%
	\BibitemOpen
	\bibfield  {author} {\bibinfo {author} {\bibfnamefont {G.}~\bibnamefont
			{Foletto}}, \bibinfo {author} {\bibfnamefont {L.}~\bibnamefont {Calderaro}},
		\bibinfo {author} {\bibfnamefont {G.}~\bibnamefont {Vallone}},\ and\ \bibinfo
		{author} {\bibfnamefont {P.}~\bibnamefont {Villoresi}},\ }\href
	{https://doi.org/10.1103/PhysRevResearch.2.033205} {\bibfield  {journal}
		{\bibinfo  {journal} {Phys. Rev. Res.}\ }\textbf {\bibinfo {volume} {2}},\
		\bibinfo {pages} {033205} (\bibinfo {year} {2020})}\BibitemShut {NoStop}%
	\bibitem [{\citenamefont {Anwer}\ \emph {et~al.}(2020)\citenamefont {Anwer},
		\citenamefont {Muhammad}, \citenamefont {Cherifi}, \citenamefont {Miklin},
		\citenamefont {Tavakoli},\ and\ \citenamefont
		{Bourennane}}]{anwer_expSQRAC_20}%
	\BibitemOpen
	\bibfield  {author} {\bibinfo {author} {\bibfnamefont {H.}~\bibnamefont
			{Anwer}}, \bibinfo {author} {\bibfnamefont {S.}~\bibnamefont {Muhammad}},
		\bibinfo {author} {\bibfnamefont {W.}~\bibnamefont {Cherifi}}, \bibinfo
		{author} {\bibfnamefont {N.}~\bibnamefont {Miklin}}, \bibinfo {author}
		{\bibfnamefont {A.}~\bibnamefont {Tavakoli}},\ and\ \bibinfo {author}
		{\bibfnamefont {M.}~\bibnamefont {Bourennane}},\ }\href
	{https://doi.org/10.1103/PhysRevLett.125.080403} {\bibfield  {journal}
		{\bibinfo  {journal} {Phys. Rev. Lett.}\ }\textbf {\bibinfo {volume} {125}},\
		\bibinfo {pages} {080403} (\bibinfo {year} {2020})}\BibitemShut {NoStop}%
	\bibitem [{\citenamefont {Xiao}\ \emph {et~al.}(2021)\citenamefont {Xiao},
		\citenamefont {Han}, \citenamefont {Fan}, \citenamefont {Qu},\ and\
		\citenamefont {Gu}}]{xiao_expSQRAC_21}%
	\BibitemOpen
	\bibfield  {author} {\bibinfo {author} {\bibfnamefont {Y.}~\bibnamefont
			{Xiao}}, \bibinfo {author} {\bibfnamefont {X.-H.}\ \bibnamefont {Han}},
		\bibinfo {author} {\bibfnamefont {X.}~\bibnamefont {Fan}}, \bibinfo {author}
		{\bibfnamefont {H.-C.}\ \bibnamefont {Qu}},\ and\ \bibinfo {author}
		{\bibfnamefont {Y.-J.}\ \bibnamefont {Gu}},\ }\href
	{https://doi.org/10.1103/PhysRevResearch.3.023081} {\bibfield  {journal}
		{\bibinfo  {journal} {Phys. Rev. Res.}\ }\textbf {\bibinfo {volume} {3}},\
		\bibinfo {pages} {023081} (\bibinfo {year} {2021})}\BibitemShut {NoStop}%
	\bibitem [{\citenamefont {Silva}\ \emph {et~al.}(2015)\citenamefont {Silva},
		\citenamefont {Gisin}, \citenamefont {Guryanova},\ and\ \citenamefont
		{Popescu}}]{silva_SequentialBell_15}%
	\BibitemOpen
	\bibfield  {author} {\bibinfo {author} {\bibfnamefont {R.}~\bibnamefont
			{Silva}}, \bibinfo {author} {\bibfnamefont {N.}~\bibnamefont {Gisin}},
		\bibinfo {author} {\bibfnamefont {Y.}~\bibnamefont {Guryanova}},\ and\
		\bibinfo {author} {\bibfnamefont {S.}~\bibnamefont {Popescu}},\ }\href
	{https://doi.org/10.1103/PhysRevLett.114.250401} {\bibfield  {journal}
		{\bibinfo  {journal} {Phys. Rev. Lett.}\ }\textbf {\bibinfo {volume} {114}},\
		\bibinfo {pages} {250401} (\bibinfo {year} {2015})}\BibitemShut {NoStop}%
	\bibitem [{\citenamefont {Mal}\ \emph {et~al.}(2016)\citenamefont {Mal},
		\citenamefont {Majumdar},\ and\ \citenamefont
		{Home}}]{mal_SequentialBell_16}%
	\BibitemOpen
	\bibfield  {author} {\bibinfo {author} {\bibfnamefont {S.}~\bibnamefont
			{Mal}}, \bibinfo {author} {\bibfnamefont {A.~S.}\ \bibnamefont {Majumdar}},\
		and\ \bibinfo {author} {\bibfnamefont {D.}~\bibnamefont {Home}},\ }\bibfield
	{journal} {\bibinfo  {journal} {Mathematics}\ }\textbf {\bibinfo {volume}
		{4}},\ \href {https://doi.org/10.3390/math4030048} {10.3390/math4030048}
	(\bibinfo {year} {2016})\BibitemShut {NoStop}%
	\bibitem [{\citenamefont {Sasmal}\ \emph {et~al.}(2018)\citenamefont {Sasmal},
		\citenamefont {Das}, \citenamefont {Mal},\ and\ \citenamefont
		{Majumdar}}]{mal_sequentialSteering_18}%
	\BibitemOpen
	\bibfield  {author} {\bibinfo {author} {\bibfnamefont {S.}~\bibnamefont
			{Sasmal}}, \bibinfo {author} {\bibfnamefont {D.}~\bibnamefont {Das}},
		\bibinfo {author} {\bibfnamefont {S.}~\bibnamefont {Mal}},\ and\ \bibinfo
		{author} {\bibfnamefont {A.~S.}\ \bibnamefont {Majumdar}},\ }\href
	{https://doi.org/10.1103/PhysRevA.98.012305} {\bibfield  {journal} {\bibinfo
			{journal} {Phys. Rev. A}\ }\textbf {\bibinfo {volume} {98}},\ \bibinfo
		{pages} {012305} (\bibinfo {year} {2018})}\BibitemShut {NoStop}%
	\bibitem [{\citenamefont {Bera}\ \emph {et~al.}(2018)\citenamefont {Bera},
		\citenamefont {Mal}, \citenamefont {Sen(De)},\ and\ \citenamefont
		{Sen}}]{mal_sequentialEntaglement_18}%
	\BibitemOpen
	\bibfield  {author} {\bibinfo {author} {\bibfnamefont {A.}~\bibnamefont
			{Bera}}, \bibinfo {author} {\bibfnamefont {S.}~\bibnamefont {Mal}}, \bibinfo
		{author} {\bibfnamefont {A.}~\bibnamefont {Sen(De)}},\ and\ \bibinfo {author}
		{\bibfnamefont {U.}~\bibnamefont {Sen}},\ }\href
	{https://doi.org/10.1103/PhysRevA.98.062304} {\bibfield  {journal} {\bibinfo
			{journal} {Phys. Rev. A}\ }\textbf {\bibinfo {volume} {98}},\ \bibinfo
		{pages} {062304} (\bibinfo {year} {2018})}\BibitemShut {NoStop}%
	\bibitem [{\citenamefont {Das}\ \emph {et~al.}(2019)\citenamefont {Das},
		\citenamefont {Ghosal}, \citenamefont {Sasmal}, \citenamefont {Mal},\ and\
		\citenamefont {Majumdar}}]{das2019facets}%
	\BibitemOpen
	\bibfield  {author} {\bibinfo {author} {\bibfnamefont {D.}~\bibnamefont
			{Das}}, \bibinfo {author} {\bibfnamefont {A.}~\bibnamefont {Ghosal}},
		\bibinfo {author} {\bibfnamefont {S.}~\bibnamefont {Sasmal}}, \bibinfo
		{author} {\bibfnamefont {S.}~\bibnamefont {Mal}},\ and\ \bibinfo {author}
		{\bibfnamefont {A.~S.}\ \bibnamefont {Majumdar}},\ }\href
	{https://doi.org/10.1103/PhysRevA.99.022305} {\bibfield  {journal} {\bibinfo
			{journal} {Phys. Rev. A}\ }\textbf {\bibinfo {volume} {99}},\ \bibinfo
		{pages} {022305} (\bibinfo {year} {2019})}\BibitemShut {NoStop}%
	\bibitem [{\citenamefont {Shenoy~H.}\ \emph {et~al.}(2019)\citenamefont
		{Shenoy~H.}, \citenamefont {Designolle}, \citenamefont {Hirsch},
		\citenamefont {Silva}, \citenamefont {Gisin},\ and\ \citenamefont
		{Brunner}}]{Akshata_19}%
	\BibitemOpen
	\bibfield  {author} {\bibinfo {author} {\bibfnamefont {A.}~\bibnamefont
			{Shenoy~H.}}, \bibinfo {author} {\bibfnamefont {S.}~\bibnamefont
			{Designolle}}, \bibinfo {author} {\bibfnamefont {F.}~\bibnamefont {Hirsch}},
		\bibinfo {author} {\bibfnamefont {R.}~\bibnamefont {Silva}}, \bibinfo
		{author} {\bibfnamefont {N.}~\bibnamefont {Gisin}},\ and\ \bibinfo {author}
		{\bibfnamefont {N.}~\bibnamefont {Brunner}},\ }\href
	{https://doi.org/10.1103/PhysRevA.99.022317} {\bibfield  {journal} {\bibinfo
			{journal} {Phys. Rev. A}\ }\textbf {\bibinfo {volume} {99}},\ \bibinfo
		{pages} {022317} (\bibinfo {year} {2019})}\BibitemShut {NoStop}%
	\bibitem [{\citenamefont {Kumari}\ and\ \citenamefont {Pan}(2019)}]{Asmita_19}%
	\BibitemOpen
	\bibfield  {author} {\bibinfo {author} {\bibfnamefont {A.}~\bibnamefont
			{Kumari}}\ and\ \bibinfo {author} {\bibfnamefont {A.~K.}\ \bibnamefont
			{Pan}},\ }\href {https://doi.org/10.1103/PhysRevA.100.062130} {\bibfield
		{journal} {\bibinfo  {journal} {Phys. Rev. A}\ }\textbf {\bibinfo {volume}
			{100}},\ \bibinfo {pages} {062130} (\bibinfo {year} {2019})}\BibitemShut
	{NoStop}%
	\bibitem [{\citenamefont {Brown}\ and\ \citenamefont
		{Colbeck}(2020)}]{brown2020arbitrarily}%
	\BibitemOpen
	\bibfield  {author} {\bibinfo {author} {\bibfnamefont {P.~J.}\ \bibnamefont
			{Brown}}\ and\ \bibinfo {author} {\bibfnamefont {R.}~\bibnamefont
			{Colbeck}},\ }\href {https://doi.org/10.1103/PhysRevLett.125.090401}
	{\bibfield  {journal} {\bibinfo  {journal} {Phys. Rev. Lett.}\ }\textbf
		{\bibinfo {volume} {125}},\ \bibinfo {pages} {090401} (\bibinfo {year}
		{2020})}\BibitemShut {NoStop}%
	\bibitem [{\citenamefont {Cheng}\ \emph {et~al.}(2021)\citenamefont {Cheng},
		\citenamefont {Liu}, \citenamefont {Baker},\ and\ \citenamefont
		{Hall}}]{cheng2021limitations}%
	\BibitemOpen
	\bibfield  {author} {\bibinfo {author} {\bibfnamefont {S.}~\bibnamefont
			{Cheng}}, \bibinfo {author} {\bibfnamefont {L.}~\bibnamefont {Liu}}, \bibinfo
		{author} {\bibfnamefont {T.~J.}\ \bibnamefont {Baker}},\ and\ \bibinfo
		{author} {\bibfnamefont {M.~J.~W.}\ \bibnamefont {Hall}},\ }\href
	{https://doi.org/10.1103/PhysRevA.104.L060201} {\bibfield  {journal}
		{\bibinfo  {journal} {Phys. Rev. A}\ }\textbf {\bibinfo {volume} {104}},\
		\bibinfo {pages} {L060201} (\bibinfo {year} {2021})}\BibitemShut {NoStop}%
	\bibitem [{\citenamefont {Roy}\ \emph {et~al.}(2024)\citenamefont {Roy},
		\citenamefont {Kumari}, \citenamefont {Mal},\ and\ \citenamefont
		{Sen(De)}}]{mal_SequentialCGLMP_24}%
	\BibitemOpen
	\bibfield  {author} {\bibinfo {author} {\bibfnamefont {S.}~\bibnamefont
			{Roy}}, \bibinfo {author} {\bibfnamefont {A.}~\bibnamefont {Kumari}},
		\bibinfo {author} {\bibfnamefont {S.}~\bibnamefont {Mal}},\ and\ \bibinfo
		{author} {\bibfnamefont {A.}~\bibnamefont {Sen(De)}},\ }\href
	{https://doi.org/10.1103/PhysRevA.109.062227} {\bibfield  {journal} {\bibinfo
			{journal} {Phys. Rev. A}\ }\textbf {\bibinfo {volume} {109}},\ \bibinfo
		{pages} {062227} (\bibinfo {year} {2024})}\BibitemShut {NoStop}%
	\bibitem [{\citenamefont {Munshi}\ and\ \citenamefont
		{Pan}(2025)}]{munshi2025device}%
	\BibitemOpen
	\bibfield  {author} {\bibinfo {author} {\bibfnamefont {S.}~\bibnamefont
			{Munshi}}\ and\ \bibinfo {author} {\bibfnamefont {A.~K.}\ \bibnamefont
			{Pan}},\ }\href {https://doi.org/10.1103/PhysRevLett.134.210203} {\bibfield
		{journal} {\bibinfo  {journal} {Phys. Rev. Lett.}\ }\textbf {\bibinfo
			{volume} {134}},\ \bibinfo {pages} {210203} (\bibinfo {year}
		{2025})}\BibitemShut {NoStop}%
	\bibitem [{\citenamefont {Curchod}\ \emph {et~al.}(2017)\citenamefont
		{Curchod}, \citenamefont {Johansson}, \citenamefont {Augusiak}, \citenamefont
		{Hoban}, \citenamefont {Wittek},\ and\ \citenamefont
		{Ac\'{\i}n}}]{Curchod_UNrand_17}%
	\BibitemOpen
	\bibfield  {author} {\bibinfo {author} {\bibfnamefont {F.~J.}\ \bibnamefont
			{Curchod}}, \bibinfo {author} {\bibfnamefont {M.}~\bibnamefont {Johansson}},
		\bibinfo {author} {\bibfnamefont {R.}~\bibnamefont {Augusiak}}, \bibinfo
		{author} {\bibfnamefont {M.~J.}\ \bibnamefont {Hoban}}, \bibinfo {author}
		{\bibfnamefont {P.}~\bibnamefont {Wittek}},\ and\ \bibinfo {author}
		{\bibfnamefont {A.}~\bibnamefont {Ac\'{\i}n}},\ }\href
	{https://doi.org/10.1103/PhysRevA.95.020102} {\bibfield  {journal} {\bibinfo
			{journal} {Phys. Rev. A}\ }\textbf {\bibinfo {volume} {95}},\ \bibinfo
		{pages} {020102} (\bibinfo {year} {2017})}\BibitemShut {NoStop}%
	\bibitem [{\citenamefont {Wagner}\ \emph {et~al.}(2020)\citenamefont {Wagner},
		\citenamefont {Bancal}, \citenamefont {Sangouard},\ and\ \citenamefont
		{Sekatski}}]{Wagner_STQInstruments_20}%
	\BibitemOpen
	\bibfield  {author} {\bibinfo {author} {\bibfnamefont {S.}~\bibnamefont
			{Wagner}}, \bibinfo {author} {\bibfnamefont {J.-D.}\ \bibnamefont {Bancal}},
		\bibinfo {author} {\bibfnamefont {N.}~\bibnamefont {Sangouard}},\ and\
		\bibinfo {author} {\bibfnamefont {P.}~\bibnamefont {Sekatski}},\ }\href
	{https://doi.org/10.22331/q-2020-03-19-243} {\bibfield  {journal} {\bibinfo
			{journal} {{Quantum}}\ }\textbf {\bibinfo {volume} {4}},\ \bibinfo {pages}
		{243} (\bibinfo {year} {2020})}\BibitemShut {NoStop}%
	\bibitem [{\citenamefont {Datta}\ and\ \citenamefont
		{Majumdar}(2018)}]{datta2018sharing}%
	\BibitemOpen
	\bibfield  {author} {\bibinfo {author} {\bibfnamefont {S.}~\bibnamefont
			{Datta}}\ and\ \bibinfo {author} {\bibfnamefont {A.~S.}\ \bibnamefont
			{Majumdar}},\ }\href {https://doi.org/10.1103/PhysRevA.98.042311} {\bibfield
		{journal} {\bibinfo  {journal} {Phys. Rev. A}\ }\textbf {\bibinfo {volume}
			{98}},\ \bibinfo {pages} {042311} (\bibinfo {year} {2018})}\BibitemShut
	{NoStop}%
	\bibitem [{\citenamefont {Maity}\ \emph {et~al.}(2020)\citenamefont {Maity},
		\citenamefont {Das}, \citenamefont {Ghosal}, \citenamefont {Roy},\ and\
		\citenamefont {Majumdar}}]{maity2020detection}%
	\BibitemOpen
	\bibfield  {author} {\bibinfo {author} {\bibfnamefont {A.~G.}\ \bibnamefont
			{Maity}}, \bibinfo {author} {\bibfnamefont {D.}~\bibnamefont {Das}}, \bibinfo
		{author} {\bibfnamefont {A.}~\bibnamefont {Ghosal}}, \bibinfo {author}
		{\bibfnamefont {A.}~\bibnamefont {Roy}},\ and\ \bibinfo {author}
		{\bibfnamefont {A.~S.}\ \bibnamefont {Majumdar}},\ }\href
	{https://doi.org/10.1103/PhysRevA.101.042340} {\bibfield  {journal} {\bibinfo
			{journal} {Phys. Rev. A}\ }\textbf {\bibinfo {volume} {101}},\ \bibinfo
		{pages} {042340} (\bibinfo {year} {2020})}\BibitemShut {NoStop}%
	\bibitem [{\citenamefont {Gupta}\ \emph {et~al.}(2021)\citenamefont {Gupta},
		\citenamefont {Maity}, \citenamefont {Das}, \citenamefont {Roy},\ and\
		\citenamefont {Majumdar}}]{gupta2021genuine}%
	\BibitemOpen
	\bibfield  {author} {\bibinfo {author} {\bibfnamefont {S.}~\bibnamefont
			{Gupta}}, \bibinfo {author} {\bibfnamefont {A.~G.}\ \bibnamefont {Maity}},
		\bibinfo {author} {\bibfnamefont {D.}~\bibnamefont {Das}}, \bibinfo {author}
		{\bibfnamefont {A.}~\bibnamefont {Roy}},\ and\ \bibinfo {author}
		{\bibfnamefont {A.~S.}\ \bibnamefont {Majumdar}},\ }\href
	{https://doi.org/10.1103/PhysRevA.103.022421} {\bibfield  {journal} {\bibinfo
			{journal} {Phys. Rev. A}\ }\textbf {\bibinfo {volume} {103}},\ \bibinfo
		{pages} {022421} (\bibinfo {year} {2021})}\BibitemShut {NoStop}%
	\bibitem [{\citenamefont {Das}\ \emph {et~al.}(2022{\natexlab{b}})\citenamefont
		{Das}, \citenamefont {Das}, \citenamefont {Mal}, \citenamefont {Home},\ and\
		\citenamefont {Majumdar}}]{das2022resource}%
	\BibitemOpen
	\bibfield  {author} {\bibinfo {author} {\bibfnamefont {A.~K.}\ \bibnamefont
			{Das}}, \bibinfo {author} {\bibfnamefont {D.}~\bibnamefont {Das}}, \bibinfo
		{author} {\bibfnamefont {S.}~\bibnamefont {Mal}}, \bibinfo {author}
		{\bibfnamefont {D.}~\bibnamefont {Home}},\ and\ \bibinfo {author}
		{\bibfnamefont {A.~S.}\ \bibnamefont {Majumdar}},\ }\href
	{https://doi.org/10.1007/s11128-022-03728-x} {\bibfield  {journal} {\bibinfo
			{journal} {Quantum Information Processing}\ }\textbf {\bibinfo {volume}
			{21}},\ \bibinfo {pages} {381} (\bibinfo {year}
		{2022}{\natexlab{b}})}\BibitemShut {NoStop}%
	\bibitem [{\citenamefont {Roy}\ \emph {et~al.}(2021)\citenamefont {Roy},
		\citenamefont {Bera}, \citenamefont {Mal}, \citenamefont {Sen(De)},\ and\
		\citenamefont {Sen}}]{roy2021recycling}%
	\BibitemOpen
	\bibfield  {author} {\bibinfo {author} {\bibfnamefont {S.}~\bibnamefont
			{Roy}}, \bibinfo {author} {\bibfnamefont {A.}~\bibnamefont {Bera}}, \bibinfo
		{author} {\bibfnamefont {S.}~\bibnamefont {Mal}}, \bibinfo {author}
		{\bibfnamefont {A.}~\bibnamefont {Sen(De)}},\ and\ \bibinfo {author}
		{\bibfnamefont {U.}~\bibnamefont {Sen}},\ }\href
	{https://doi.org/https://doi.org/10.1016/j.physleta.2021.127143} {\bibfield
		{journal} {\bibinfo  {journal} {Physics Letters A}\ }\textbf {\bibinfo
			{volume} {392}},\ \bibinfo {pages} {127143} (\bibinfo {year}
		{2021})}\BibitemShut {NoStop}%
	\bibitem [{\citenamefont {Datta}\ \emph {et~al.}(2024)\citenamefont {Datta},
		\citenamefont {Mal}, \citenamefont {Pati},\ and\ \citenamefont
		{Majumdar}}]{datta2024remote}%
	\BibitemOpen
	\bibfield  {author} {\bibinfo {author} {\bibfnamefont {S.}~\bibnamefont
			{Datta}}, \bibinfo {author} {\bibfnamefont {S.}~\bibnamefont {Mal}}, \bibinfo
		{author} {\bibfnamefont {A.~K.}\ \bibnamefont {Pati}},\ and\ \bibinfo
		{author} {\bibfnamefont {A.~S.}\ \bibnamefont {Majumdar}},\ }\href
	{https://doi.org/10.1007/s11128-024-04263-7} {\bibfield  {journal} {\bibinfo
			{journal} {Quantum Information Processing}\ }\textbf {\bibinfo {volume}
			{23}},\ \bibinfo {pages} {54} (\bibinfo {year} {2024})}\BibitemShut {NoStop}%
	\bibitem [{\citenamefont {Mohan}\ \emph {et~al.}(2019)\citenamefont {Mohan},
		\citenamefont {Tavakoli},\ and\ \citenamefont {Brunner}}]{Mohan_STI19}%
	\BibitemOpen
	\bibfield  {author} {\bibinfo {author} {\bibfnamefont {K.}~\bibnamefont
			{Mohan}}, \bibinfo {author} {\bibfnamefont {A.}~\bibnamefont {Tavakoli}},\
		and\ \bibinfo {author} {\bibfnamefont {N.}~\bibnamefont {Brunner}},\ }\href
	{https://doi.org/10.1088/1367-2630/ab3773} {\bibfield  {journal} {\bibinfo
			{journal} {New Journal of Physics}\ }\textbf {\bibinfo {volume} {21}},\
		\bibinfo {pages} {083034} (\bibinfo {year} {2019})}\BibitemShut {NoStop}%
	\bibitem [{\citenamefont {Miklin}\ \emph {et~al.}(2020)\citenamefont {Miklin},
		\citenamefont {Borka\l{}a},\ and\ \citenamefont
		{Paw\l{}owski}}]{Miklin_SDIM20}%
	\BibitemOpen
	\bibfield  {author} {\bibinfo {author} {\bibfnamefont {N.}~\bibnamefont
			{Miklin}}, \bibinfo {author} {\bibfnamefont {J.~J.}\ \bibnamefont
			{Borka\l{}a}},\ and\ \bibinfo {author} {\bibfnamefont {M.}~\bibnamefont
			{Paw\l{}owski}},\ }\href {https://doi.org/10.1103/PhysRevResearch.2.033014}
	{\bibfield  {journal} {\bibinfo  {journal} {Phys. Rev. Res.}\ }\textbf
		{\bibinfo {volume} {2}},\ \bibinfo {pages} {033014} (\bibinfo {year}
		{2020})}\BibitemShut {NoStop}%
	\bibitem [{\citenamefont {Tavakoli}\ \emph {et~al.}(2020)\citenamefont
		{Tavakoli}, \citenamefont {Smania}, \citenamefont {Vértesi}, \citenamefont
		{Brunner},\ and\ \citenamefont {Bourennane}}]{Tavakoli_STM20}%
	\BibitemOpen
	\bibfield  {author} {\bibinfo {author} {\bibfnamefont {A.}~\bibnamefont
			{Tavakoli}}, \bibinfo {author} {\bibfnamefont {M.}~\bibnamefont {Smania}},
		\bibinfo {author} {\bibfnamefont {T.}~\bibnamefont {Vértesi}}, \bibinfo
		{author} {\bibfnamefont {N.}~\bibnamefont {Brunner}},\ and\ \bibinfo {author}
		{\bibfnamefont {M.}~\bibnamefont {Bourennane}},\ }\href
	{https://doi.org/10.1126/sciadv.aaw6664} {\bibfield  {journal} {\bibinfo
			{journal} {Science Advances}\ }\textbf {\bibinfo {volume} {6}},\ \bibinfo
		{pages} {eaaw6664} (\bibinfo {year} {2020})}\BibitemShut {NoStop}%
	\bibitem [{\citenamefont {Das}\ \emph {et~al.}(2021)\citenamefont {Das},
		\citenamefont {Ghosal}, \citenamefont {Maity}, \citenamefont {Kanjilal},\
		and\ \citenamefont {Roy}}]{debarshi_seq_RAC24}%
	\BibitemOpen
	\bibfield  {author} {\bibinfo {author} {\bibfnamefont {D.}~\bibnamefont
			{Das}}, \bibinfo {author} {\bibfnamefont {A.}~\bibnamefont {Ghosal}},
		\bibinfo {author} {\bibfnamefont {A.~G.}\ \bibnamefont {Maity}}, \bibinfo
		{author} {\bibfnamefont {S.}~\bibnamefont {Kanjilal}},\ and\ \bibinfo
		{author} {\bibfnamefont {A.}~\bibnamefont {Roy}},\ }\href
	{https://doi.org/10.1103/PhysRevA.104.L060602} {\bibfield  {journal}
		{\bibinfo  {journal} {Phys. Rev. A}\ }\textbf {\bibinfo {volume} {104}},\
		\bibinfo {pages} {L060602} (\bibinfo {year} {2021})}\BibitemShut {NoStop}%
	\bibitem [{\citenamefont {Schiavon}\ \emph {et~al.}(2017)\citenamefont
		{Schiavon}, \citenamefont {Calderaro}, \citenamefont {Pittaluga},
		\citenamefont {Vallone},\ and\ \citenamefont
		{Villoresi}}]{schiavon2017three}%
	\BibitemOpen
	\bibfield  {author} {\bibinfo {author} {\bibfnamefont {M.}~\bibnamefont
			{Schiavon}}, \bibinfo {author} {\bibfnamefont {L.}~\bibnamefont {Calderaro}},
		\bibinfo {author} {\bibfnamefont {M.}~\bibnamefont {Pittaluga}}, \bibinfo
		{author} {\bibfnamefont {G.}~\bibnamefont {Vallone}},\ and\ \bibinfo {author}
		{\bibfnamefont {P.}~\bibnamefont {Villoresi}},\ }\href
	{https://doi.org/10.1088/2058-9565/aa62be} {\bibfield  {journal} {\bibinfo
			{journal} {Quantum Science and Technology}\ }\textbf {\bibinfo {volume}
			{2}},\ \bibinfo {pages} {015010} (\bibinfo {year} {2017})}\BibitemShut
	{NoStop}%
	\bibitem [{\citenamefont {Hu}\ \emph {et~al.}(2018)\citenamefont {Hu},
		\citenamefont {Zhou}, \citenamefont {Hu}, \citenamefont {Li}, \citenamefont
		{Guo},\ and\ \citenamefont {Zhang}}]{hu2018observation}%
	\BibitemOpen
	\bibfield  {author} {\bibinfo {author} {\bibfnamefont {M.-J.}\ \bibnamefont
			{Hu}}, \bibinfo {author} {\bibfnamefont {Z.-Y.}\ \bibnamefont {Zhou}},
		\bibinfo {author} {\bibfnamefont {X.-M.}\ \bibnamefont {Hu}}, \bibinfo
		{author} {\bibfnamefont {C.-F.}\ \bibnamefont {Li}}, \bibinfo {author}
		{\bibfnamefont {G.-C.}\ \bibnamefont {Guo}},\ and\ \bibinfo {author}
		{\bibfnamefont {Y.-S.}\ \bibnamefont {Zhang}},\ }\href
	{https://doi.org/10.1038/s41534-018-0115-x} {\bibfield  {journal} {\bibinfo
			{journal} {npj Quantum Information}\ }\textbf {\bibinfo {volume} {4}},\
		\bibinfo {pages} {63} (\bibinfo {year} {2018})}\BibitemShut {NoStop}%
	\bibitem [{\citenamefont {Ambainis}\ \emph {et~al.}(2024)\citenamefont
		{Ambainis}, \citenamefont {Kravchenko}, \citenamefont {Sazim}, \citenamefont
		{Bae},\ and\ \citenamefont {Rai}}]{Ambainis_QRAC24}%
	\BibitemOpen
	\bibfield  {author} {\bibinfo {author} {\bibfnamefont {A.}~\bibnamefont
			{Ambainis}}, \bibinfo {author} {\bibfnamefont {D.}~\bibnamefont
			{Kravchenko}}, \bibinfo {author} {\bibfnamefont {S.}~\bibnamefont {Sazim}},
		\bibinfo {author} {\bibfnamefont {J.}~\bibnamefont {Bae}},\ and\ \bibinfo
		{author} {\bibfnamefont {A.}~\bibnamefont {Rai}},\ }\href
	{https://doi.org/10.1088/1367-2630/ad9bdf} {\bibfield  {journal} {\bibinfo
			{journal} {New Journal of Physics}\ }\textbf {\bibinfo {volume} {26}},\
		\bibinfo {pages} {123023} (\bibinfo {year} {2024})}\BibitemShut {NoStop}%
	\bibitem [{\citenamefont {Busch}(1986)}]{Busch1986}%
	\BibitemOpen
	\bibfield  {author} {\bibinfo {author} {\bibfnamefont {P.}~\bibnamefont
			{Busch}},\ }\href {https://doi.org/10.1103/PhysRevD.33.2253} {\bibfield
		{journal} {\bibinfo  {journal} {Phys. Rev. D}\ }\textbf {\bibinfo {volume}
			{33}},\ \bibinfo {pages} {2253} (\bibinfo {year} {1986})}\BibitemShut
	{NoStop}%
	\bibitem [{\citenamefont {Liabtro}(2017)}]{liabotro_ImprovedBnd_17}%
	\BibitemOpen
	\bibfield  {author} {\bibinfo {author} {\bibfnamefont {O.}~\bibnamefont
			{Liabtro}},\ }\href {https://doi.org/10.1103/PhysRevA.95.052315} {\bibfield
		{journal} {\bibinfo  {journal} {Phys. Rev. A}\ }\textbf {\bibinfo {volume}
			{95}},\ \bibinfo {pages} {052315} (\bibinfo {year} {2017})}\BibitemShut
	{NoStop}%
	\bibitem [{\citenamefont {Werner}\ and\ \citenamefont
		{Wolf}(2001)}]{wernerwolf_01}%
	\BibitemOpen
	\bibfield  {author} {\bibinfo {author} {\bibfnamefont {R.~F.}\ \bibnamefont
			{Werner}}\ and\ \bibinfo {author} {\bibfnamefont {M.~M.}\ \bibnamefont
			{Wolf}},\ }\href {https://dl.acm.org/doi/abs/10.5555/2011339.2011340}
	{\bibfield  {journal} {\bibinfo  {journal} {Quantum Info. Comput.}\ }\textbf
		{\bibinfo {volume} {1}},\ \bibinfo {pages} {1–25} (\bibinfo {year}
		{2001})}\BibitemShut {NoStop}%
\end{thebibliography}
%


\appendix
\onecolumngrid

\newpage

\section{Optimal of success probability and self-testing}
\label{proof_lemma}
The success probability of the second receiver, $P_{AC}^{Q}$ is given by,
\begin{equation}
    \begin{aligned}
        P_{AC}^{Q} = \frac{1}{8} \sum_{x,y,b} \tr(K_{b|y}\rho_{x}K_{b|y}^{\dagger} N_{x_{\bar{y}}|\bar{y}})
    \end{aligned}
\end{equation}
under the constraint that the success probability of the first receiver
is,
\begin{equation}
    \begin{aligned}
        \beta = \frac{1}{8} \sum_{x,y} \tr(\rho_{x}K_{b|y}^{\dagger} K_{b|y})
    \end{aligned}
    \label{A2}
\end{equation}
We can use polar decomposition $K_{b|y} = U_{yb}\sqrt{M_{b|y}}$ and substitute $N_{1|z} = \mathbb{I} - N_{0|z}$, to get

\begin{equation}
    \begin{aligned}
        P_{AC}^{Q} = \frac{1}{2} +\frac{1}{8} \sum_{x,y,b}(-1)^{\bar{y}} \tr(\sqrt{M_{b|y}}\rho_{x}\sqrt{M_{b|y}} U_{yb}^{\dagger} N_{0|\bar{y}}U_{yb})
    \end{aligned}
\end{equation}
Taking $\gamma_{y} = \sum_{x}(-1)^{x_{\bar{y}}}\rho_x$ and $D_{y b}^{\bar{y}} = U_{yb}^{\dagger} N_{x_{\bar{y}}|\bar{y}}U_{yb}$, we get

\begin{equation}
    \begin{aligned}
        \max P_{AC}^{Q} &= \frac{1}{2} + \max_{\rho,D,M} \frac{1}{8}\sum_{y,b}\tr(\sqrt{M_{b|y}}\gamma_{z}\sqrt{M_{b|y}}D_{y b}^{\bar{y}})\\
        &=\frac{1}{2} + \max_{\rho,D,M} \frac{1}{8}\sum_{y,b}\lambda_{max}(\sqrt{M_{b|y}}\gamma_{z}\sqrt{M_{b|y}})
    \end{aligned}
    \label{optPAC}
\end{equation}
and
\begin{equation}
    \begin{aligned}
        \gamma_{y} = \frac{1}{2}\left[ (\eta_{00}-\eta_{11}) + (-1)^{\bar{y}}(\eta_{01}-\eta_{10})\right]\cdot\Vec{\sigma}
     \end{aligned}
\end{equation}
Let $\Vec{q}_{\bar{y}} = \frac{1}{2}\left[ (\eta_{00}-\eta_{11}) + (-1)^{\bar{y}}(\eta_{01}-\eta_{10})\right]$. Now, 
\begin{equation}
    \begin{aligned}
        \forall M, \forall \Vec{v} \in \mathbb{R}^{3}: \quad \sum_{b = 0,1} \lambda_{max}[\sqrt{M_{b}}(\Vec{v}\cdot \Vec{\sigma})\sqrt{M_{b}}]\leq |\Vec{v}|
    \end{aligned}
\end{equation}
and
\begin{equation}
    \begin{aligned}
        P_{AC}^{Q} \leq \frac{1}{2} + \frac{1}{8}\sum_{y,b}\lambda_{\max}\left[\sqrt{M_{b|y}}(\Vec{q}_{\bar{y}}\cdot\Vec{\sigma})\sqrt{M_{b|y}}\right]
    \end{aligned}
\end{equation}
Now define $T \equiv \sum_{y,b}\lambda_{\max}\left[\sqrt{M_{b|y}}(\Vec{q}_{\bar{y}}\cdot\Vec{\sigma})\sqrt{M_{b|y}}\right]$.
Let's consider an observable $M_{y} = M_{0|y}-M_{1|y}$. We can write $M_{y} = g_{y0} \mathbb{I} + \Vec{g}_{y}\cdot\Vec{\sigma}$ where $\Vec{g}_{y} = (g_{y1},g_{y2},g_{y3})$ with $|\Vec{g}_{y}|\leq 1$ and $|\Vec{g}_{y}|-1 \leq g_{y0} \leq 1-|\Vec{g}_{y}|$. This constraint ensures the operator is positive:
\begin{equation}
    \begin{aligned}
        M_{b|y} = f_{yb}\ket{\Vec{g}_{y}}\bra{\Vec{g}_{y}}+h_{yb}\ket{-\Vec{g}_{y}}\bra{-\Vec{g}_{y}}
    \end{aligned}
\end{equation}
Here, $\ket{\Vec{g}_{y}}$ is the pure state corresponding to the bloch sphere direction $\Vec{g}_{y}$,
\begin{equation}
    \begin{aligned}
        f_{yb} = \frac{1}{2}(1+(-1)^b g_{y0} +(-1)^b |\Vec{g_{y}}|) \nonumber \\
        h_{yb} = \frac{1}{2}(1+(-1)^b g_{y0} -(-1)^b |\Vec{g_{y}}|) \nonumber \\
    \end{aligned}
\end{equation}
Now, we can write Eq.~(\ref{A2}) as,
\begin{equation}
    \begin{aligned}
        P_{AB}^{Q} = \beta = \frac{1}{8}(4+|\Vec{q_0}|g_{01}+|\Vec{q_1}|g_{13})
    \end{aligned}
    \label{PABQB}
\end{equation}
Then,  solving the characteristic equation $\text{det}\left(\sqrt{M_{b|y}}(\Vec{q}_{\Bar{y}}\cdot\Vec{\sigma})\sqrt{M_{b|y}} - \nu \mathbb{I}\right) = 0$, we get,
\begin{equation}
    \begin{aligned}
        T = \sum_{y,b} \frac{|\Vec{q}_{\bar{y}}|}{2}\sqrt{(1+(-1)^b g_{y0})^2 -|\Vec{g}_{y}|^2 (1 - \bra{\Vec{g}_{y}}\hat{q}_{\bar{y}}\cdot\Vec{\sigma}\ket{\Vec{g}_{y}}^2)}
    \end{aligned}
\end{equation}

We now perform the optimization over $g_{y0}$ by  considering the two terms corresponding to $y = 0, 1$ separately.  This amounts maximizing an expression of the form $\sqrt{(1+w)^2 - z}+\sqrt{(1-w)^2 - z}$, for  positive $z$. It can be shown such that such a function is uniquely maximized by setting $w=0$. So we require $g_{00} = g_{10} =0$. Moreover, $(\Vec{q}_{0}, \Vec{q}_{1})$ has no component along y-axis, and one can optimally choose $g_{02} = g_{12} =0$. This enables us to express 
\begin{equation}\label{eqT}
    \begin{aligned}
        \max T = |\Vec{q}_{0}|\sqrt{1-(g_{11}^2+g_{13}^2)(1-g_{11}^2)}+|\Vec{q}_{1}|\sqrt{1-(g_{01}^2+g_{03}^2)(1-g_{03}^2)}
    \end{aligned}
\end{equation}
Note that $g_{03}$ and $g_{11}$ do not appear in the constraint, and that they are associated with different settings of Barun,  and  appear in different terms in Eq.~(\ref{eqT}). Thus we can separately maximize using standard differentiation. This returns  the unique maximum  attained for $g_{03} = g_{11} = 0$, and hence, we have,
\begin{equation}
    \begin{aligned}
        P_{AC}^{Q} \leq \frac{1}{2}+\frac{1}{8}(|\Vec{q}_{0}|\sqrt{1-g_{13}^{2}}+|\Vec{q}_{1}|\sqrt{1-g_{01}^{2}}) \equiv \Tilde{P}
    \end{aligned}
\end{equation}
Let $g_{01} = \cos\zeta_0$ and $g_{13} = \cos\zeta_1$ for $\zeta_{0},\zeta_{1}\in [0,\frac{\pi}{2}]$, so that
\begin{equation}
    \begin{aligned}
        \Tilde{P} = \frac{1}{2}+\frac{1}{4}(\cos\frac{\theta}{2}\sin\zeta_1+\sin\frac{\theta}{2}\sin\zeta_0)
    \end{aligned}
    \label{Ptilde}
\end{equation}
By putting the relevant values we can rewrite $\beta$ as,
\begin{equation}
    \begin{aligned}
        \beta = \frac{1}{8}(4+\cos\frac{\theta}{2}\cos\zeta_0+\sin\frac{\theta}{2}\cos\zeta_1)
    \end{aligned}
    \label{beta}
\end{equation}


\textbf{Lemma 1:} For every tuple $(\theta, \zeta_0, \zeta_1)$ corresponding to $(\beta, \Tilde{P})$, there exists another tuple $(\theta, \zeta_0,\zeta_1) = (\frac{\pi}{2}, \zeta,\zeta)$ that produces $(\beta, \Tilde{P}^{'})$ with $\Tilde{P}^{'}\geq \Tilde{P}$, Moreover, $\theta = \frac{\pi}{2}$ and $\zeta_0 = \zeta_1$ leads to the optimal $\Tilde{P}^{'}$.

\textbf{Proof:}

To prove this lemma, we must show that $\forall \theta, \zeta_0, \zeta_1 \in [0,\frac{\pi}{2}]$ there exists $\zeta \in [0,\frac{\pi}{2}]$ such that,
\begin{equation}
    \begin{aligned}
        \cos\frac{\theta}{2}\cos\zeta_0 + \sin\frac{\theta}{2}\cos\zeta_1 = \sqrt{2}\cos\zeta \nonumber \\
        \cos\frac{\theta}{2}\sin\zeta_0 + \sin\frac{\theta}{2}\sin\zeta_1 \leq \sqrt{2}\sin\zeta
    \end{aligned}
    \label{comb}
\end{equation}
By squaring both equations and inequalities of Eq.~(\ref{comb}), we can combine them to eliminate $\zeta$: 
\begin{equation}
    \begin{aligned}
        \cos\theta (\cos^2 \zeta_0 - \cos^2 \zeta_1) + \sin\theta \cos(\zeta_0-\zeta_1)  \leq 1
    \end{aligned}
\end{equation}
Differentiating w.r.t. $\zeta_0$, one finds that the optimum of the left-hand side is attained for $\zeta_1 = \zeta_0$, which proves the above relation.

From the lemma 1, we set $\theta = \frac{\pi}{2}$ in (\ref{Ptilde}) and (\ref{beta}), and $g_{01} = g_{13} = g$. Then,
\begin{equation}
    \begin{aligned}
        g = \sqrt{2}(2 \beta - 1).
    \end{aligned}
    \label{g}
\end{equation}
We also have $\Tilde{P} = \frac{1}{2} + \frac{\sqrt{1-g^2}}{2\sqrt{2}}$, and
hence,
\begin{equation}
    \begin{aligned}
        P_{AC}^{Q,\beta} \leq \frac{1}{4}(2+\sqrt{16\beta -16\beta^2 -2}).
    \end{aligned}
    \label{upbound}
\end{equation}
 The r.h.s. of Eq.(\ref{upbound}) corresponds to the success probabilities
 emanating from the quantum strategy (\ref{tradeoff}) in the
 main text, showing that the above upper bound could be saturated with an explicit quantum strategy.

In order to analyze the self-testing argument explicitly, let us more closely examine the  steps needed to arrive at Eq.~ (\ref{upbound}). First, we have already shown that the preparations must be pure, pairwise antipodal, and by lemma 1, they must have a relative angle of $\frac{\pi}{2}$. This corresponds to a square in a disk of the Bloch sphere. The above arguments fully characterize Aparna’s preparations up to a reference frame. 
For Barun’s instrument, we have shown that the Bloch vectors ($\vec{g}_0, \vec{g}_1$) can only have non-zero components in the $x$- and $z$-directions respectively, and that the length of the Bloch vector is given by Eq.~(\ref{g}). 
This fully characterizes the Bloch vectors.
Moreover, in Eq.~(\ref{optPAC}), we require that $D_{y b}^{\bar{y}}$ is aligned 
with the eigenvector of $\sqrt{M_{b|y}} D_{y b} \sqrt{M_{b|y}}$
corresponding to the largest eigenvalue. In case of Chhanda's measurement, we now have 
$\gamma_0 = \sigma_x$ and $\gamma_1 = \sigma_z$, while $M_0 \propto \sigma_x$ 
and $M_1 \propto \sigma_z$. 
Therefore, it follows that $\forall y, b : D_{y b}^0 = |+\rangle\langle+|$ 
and $\forall y, b : D_{y b}^1 = |0\rangle\langle0|$,  we have,
\begin{equation}
    \begin{aligned}
        U_{y b}^{\dagger}N_{0|0}U_{y b} = \ket{+}\bra{+}\\
        U_{y b}^{\dagger}N_{0|1}U_{y b} = \ket{0}\bra{0}
    \end{aligned}
\end{equation}
This implies that all unitaries are equal: $U_{y b} = U$. 
Therefore, Chhanda’s observables $C_z = C_{0|z} - C_{1|z}$ satisfy 
$C_0 = U \sigma_x U^\dagger \quad \text{and} \quad C_1 = U \sigma_z U^\dagger.$ This completes our self-testing argument

\section{Comparison of bound gaps for different visibility settings}
\label{ule}

\begin{figure*}[ht]
    \centering

    \begin{subfigure}{0.48\textwidth}
        \centering
        \includegraphics[width=\linewidth]{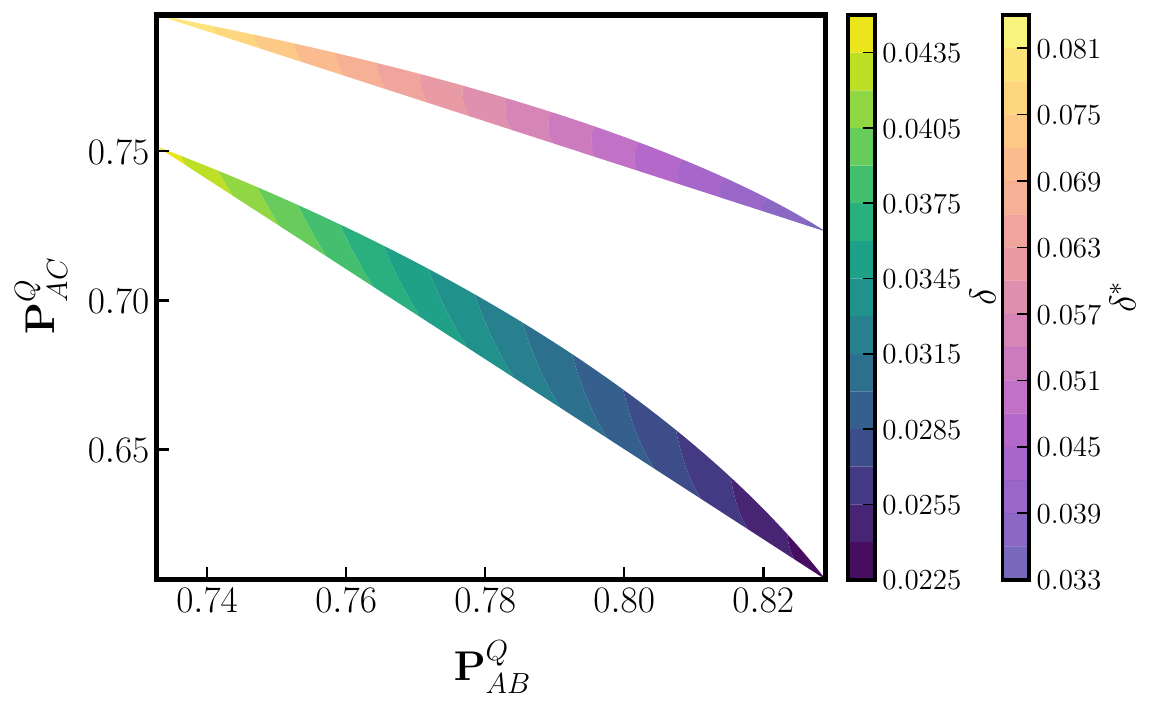}
        \caption{}
        \label{Plot2}
    \end{subfigure}
    \hfill
    \begin{subfigure}{0.48\textwidth}
        \centering
        \includegraphics[width=\linewidth]{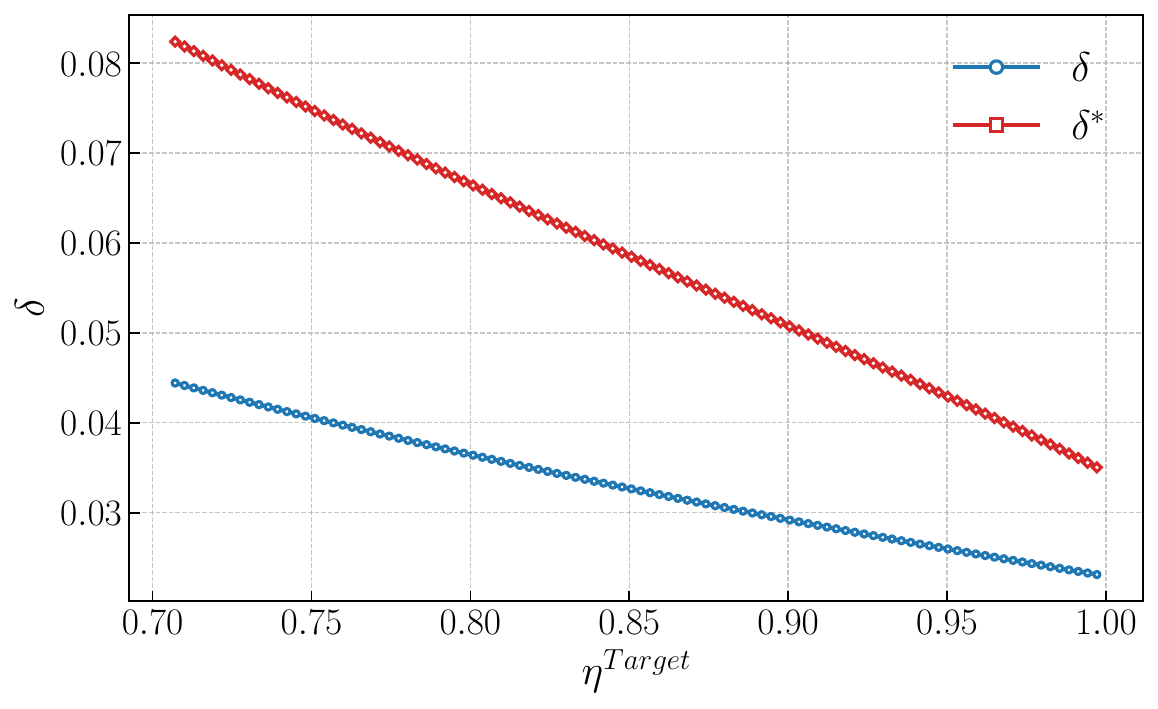}
        \caption{}
        \label{Plot2.1}
    \end{subfigure}

    \vspace{0.5cm}

    \begin{subfigure}{0.48\textwidth}
        \centering
        \includegraphics[width=\linewidth]{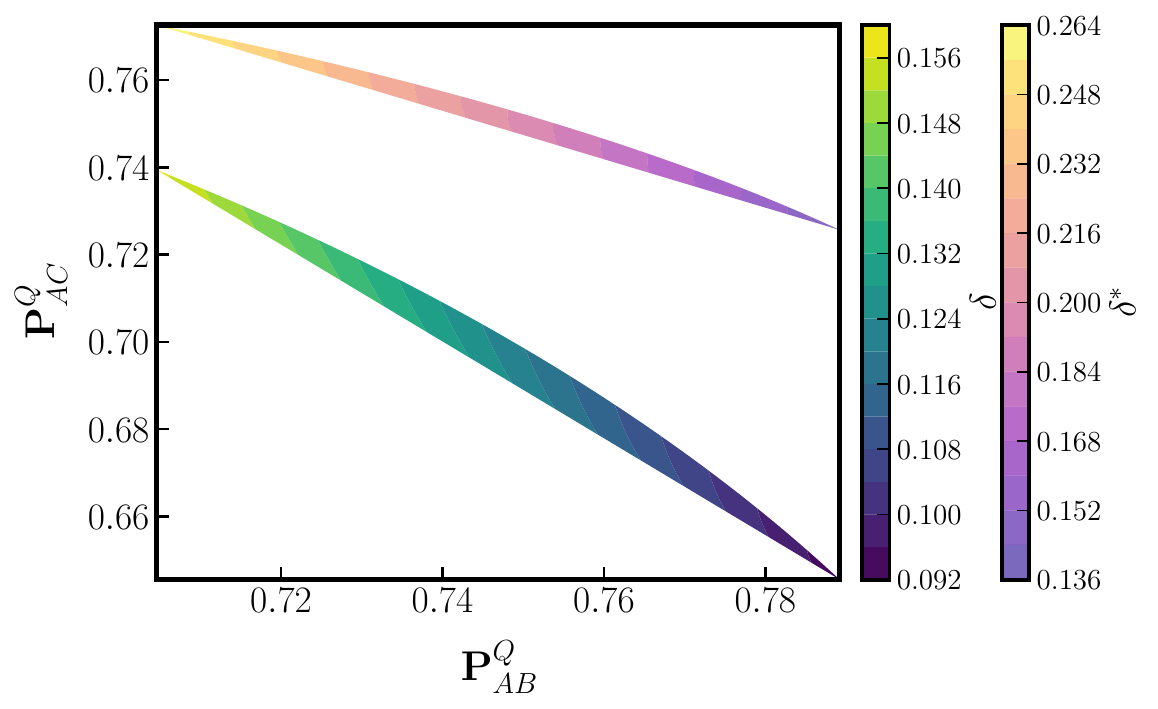}
        \caption{}
        \label{Plot3}
    \end{subfigure}
    \hfill
    \begin{subfigure}{0.48\textwidth}
        \centering
        \includegraphics[width=\linewidth]{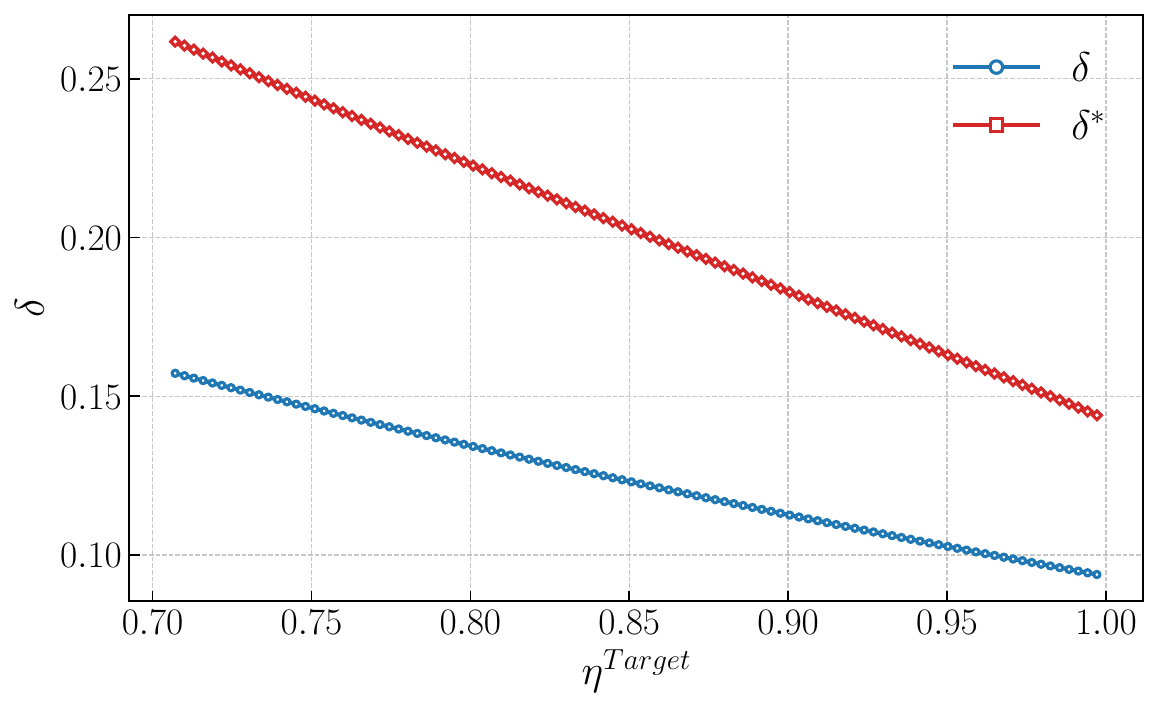}
        \caption{}
        \label{Plot3.1}
    \end{subfigure}

    \caption{
        (a) Contour plots of $\delta$ and $\delta^*$ as functions of $\text{P}_{\text{AB}}^Q$ and $\text{P}_{\text{AC}}^Q$ for $p_1 = 0.98$, $p_2 = 0.95$, $p_3 = 0.98$.  
(b) Plots of $\delta$ and $\delta^*$ as functions of $\eta^{\text{Target}}$ for the same visibility values.  
(c) Contour plots of $\delta$ and $\delta^*$ as functions of $\text{P}_{\text{AB}}^Q$ and $\text{P}_{\text{AC}}^Q$ for $p_1 = 0.93$, $p_2 = 0.88$, $p_3 = 0.93$.  
(d) Plots of $\delta$ and $\delta^*$ as functions of $\eta^{\text{Target}}$ for the same visibility values. 
Here, $\delta$ denotes the gap between the upper and lower bounds of the sharpness parameter estimated from the multi-receiver communication game proposed in this work, while $\delta^*$ represents the corresponding bound-gap reported in Ref.~\cite{Mohan_STI19}.
 }
    \label{Comp_graphs_APP}
\end{figure*}


\section{Success probabilities for higher-dimensional systems}

\label{SEE-SAW-Algo}

Here we employ the technique of the  see-saw algorithm, which 
  is especially well-suited to our optimization problem required to generalize the analysis in Section~\ref{QG} to higher dimensional systems. 
  Our goal here is to maximize the success probabilities over $d$-dimensional quantum states $\rho_x$ and $d$- outcome POVMs. The success probability of the first receiver is obtained as,

\begin{equation}
\begin{aligned}
& \text{max} \; P_{AB}^{Q} = \frac{1}{2 d^2} \sum_{x, y} Tr(\rho_x M_{b=x_{y}|y})\\
\text{subject to} \quad 
&\forall x : \rho_x \in \mathbb{C}^d, \; \rho_x \geq 0, \; \text{Tr}(\rho_x) = 1,\\
&\forall y, b:  \; M_{b|y} \geq 0, \; \sum_{b = 0}^{d-1} M_{b|y}= \mathbb{I}.
\label{opt1}
\end{aligned}
\end{equation}
The objective function in the above maximization problem~(\ref{opt1}), is a product of semi-definite matrices. The see-saw algorithm is an iterative optimization technique that alternates between fixing one matrix and optimizing over the other. For the sake of completeness let's describe the algorithm briefly by considering the maximization of $P_{AB}^Q$. First, we randomly select an initial state for Aparna's preparation and proceed to optimize the objective function over Barun's POVMs.

\begin{equation}
\begin{aligned}
& \text{max} \; P_{AB}^{Q} = \frac{1}{2 d^2} \sum_{x, y} Tr(\boldsymbol{\rho_x} M_{b=x_{y}|y})\\
\text{s. t.} \quad 
&\forall y, b:  \; M_{b|y} \geq 0, \; \sum_{b = 0}^{d-1} M_{b|y}= \mathbb{I}.
\end{aligned}
\label{opt2}
\end{equation}
Symbols in bold indicates which variable is kept fixed.
Since the objective function is linear in the semi-definite variables, the problem is an SDP. In the second step we fix the optimal measurements for Barun, obtained from the previous SDP, and optimize over Aparna's preparations, i.e., 
\begin{equation}
\begin{aligned}
& \text{max} \; P_{AB}^{Q} = \frac{1}{2 d^2} \sum_{x, y} Tr(\rho_x \boldsymbol{M_{b=x_{y}|y}})\\
\text{s. t.} \quad 
&\forall x : \rho_x \in \mathbb{C}^d, \; \rho_x \geq 0,\; \text{Tr}(\rho_x) = 1.
\end{aligned}
\label{opt3}
\end{equation}
 This problem is again an SDP. Now in the first step of the next iteration, Aparna's optimal state preparation, coming out of the last optimization problem, is kept fixed and the SDP (\ref{opt2}) is run followed by (\ref{opt3}). In this way the algorithm is run in a loop several times until it converges to an apriori fixed value of precision. While this approach often yields good results, converge to a global maximum is not guaranteed.  To increase the likelihood of finding a global maximum, the entire procedure is repeated multiple times with different randomly chosen initial states. 

 We can apply the above stated see-saw technique to optimize the total success probability of the proposed communication game. Here, Barun performs an unsharp measurement of the form $\Tilde{M}_{b|y} = \eta M_{b|y} + (1-\eta)\frac{\mathbb{I}}{d}$ with sharpness parameter $\eta \in [0,1]$. A natural choice for the Kraus operator leads to Lüders' instrument \cite{Busch1986}, which was proven to be optimal in the context of measurement-information gain trade-off \cite{mal_SequentialBell_16}, is given as,
\begin{equation}
\label{Kraus}
\begin{aligned}
K_{b|y} = \left(\sqrt{\frac{1+(d-1)\eta}{d}}-\sqrt{\frac{1-\eta}{d}}\right)M_{b|y}+\sqrt{\frac{1-\eta}{d}}\mathbb{I}.
\end{aligned}
\end{equation}
For this unsharp measurement, for a given value of $\eta$, the success probability of Barun and subsequently that of Chhanda can be computed using the see-saw method to obtain a lower bound on the optimal value.


Let's describe the relevant steps employing the see-saw algorithm to compute the total success probability of our two-receiver communication game with restricted collaboration. 

\textit{Step 1:}
We initialize Aparna's state preparation, $\{\boldsymbol{\rho_x}\}$, with random quantum states and proceed to maximize the objective function for a given value of \(\eta\) over Barun's measurements. 
\begin{equation}
\begin{aligned}
& \text{max} \; P_{AB}^{Q} = \frac{1}{2 d^2} \sum_{x, y} Tr(\boldsymbol{\rho_x} \tilde{M}_{b=x_{y}|y})\\
\text{subject to} \quad 
&\forall y, b:  \; \tilde{M}_{b|y} \geq 0, \; \sum_{b = 0}^{d-1} \tilde{M}_{b|y}= \mathbb{I}.
\end{aligned}
\label{optS1}
\end{equation}

\textit{Step 2:}
Then apply the Kraus operators as Eq.~(\ref{Kraus}) on the randomly chosen states in the previous step and maximize Chhanda's success over her POVM.

\begin{equation}
\begin{aligned}
& \text{max} \; P_{AC}^{Q} = \frac{1}{2 d^2} \sum_{x, b, \bar{y}} \text{Tr}(K_{b|y}\boldsymbol{\rho_x} K_{b|y}^{\dagger} N_{x_{\bar{y}}|\bar{y}})\\
\text{subject to} \quad 
&\forall y, b:  \; N_{b|\bar{y}} \geq 0, \; \sum_{b = 0}^{d-1} N_{b|\bar{y}}= \mathbb{I}.
\end{aligned}
\label{optS2}
\end{equation}

\textit{Step 3:}
We fix the optimal POVMS obtained from step 1 as Barun's measurements (\(\boldsymbol{\tilde{M}_{b=x_{y}|y}}\)) and optimize Aparna's preparations to maximize the objective function.
\begin{equation}
\begin{aligned}
& \text{max} \; P_{AB}^{Q} = \frac{1}{2 d^2} \sum_{x, y} Tr(\rho_x \boldsymbol{\tilde{M}_{b=x_{y}|y}})\\
\text{subject to} \quad 
&\forall x : \rho_x \in \mathbb{C}^d, \; \rho_x \geq 0,\; \text{Tr}(\rho_x) = 1.
\end{aligned}
\label{optS3}
\end{equation}
These three steps are repeated iteratively until convergence in both \(P_{AB}^{Q}\) and \(P_{AC}^{Q}\) is achieved.

\end{document}